\documentclass[12pt]{article}
\usepackage{a4wide}
\usepackage{graphicx}
\usepackage{amssymb}
\usepackage{amsmath}
\usepackage{slashed}
\usepackage{multicol}
\usepackage{braket}
\newcommand{\be}{\begin{equation}}
\newcommand{\ee}{\end{equation}}
\newcommand{\ba}{\begin{eqnarray}}
\newcommand{\ea}{\end{eqnarray}}

\begin{document}

\allowdisplaybreaks

\begin{titlepage}
\begin{flushright}
\end{flushright}
\vfill
\begin{center}
{\Large\bf Split fermionic WIMPs evade direct detection}
\vfill
{\bf Karim Ghorbani \footnote{karim1.ghorbani@gmail.com} }\\[1cm]
{Physics Department, Faculty of Sciences, Arak University, Arak 38156-8-8349, Iran}\\[.4cm]

\end{center}

\vfill

\begin{abstract}
We consider a model with two gauge singlet fermionic WIMPs communicating with the SM particles by a 
singlet scalar mediator via a Higgs portal. 
While the light WIMP is stable and plays the role of the dark matter (DM) candidate, the heavy partner is a short-lived WIMP without
contribution to the current DM relic density. Along with the coannihilation effects the heavy WIMP, acting as a mediator 
in $t$- and $u$-channel DM annihilation cross sections, has a significant effect in finding the viable parameter 
space against the direct detection constraints provided by XENON1t and LUX experiments.
This is an extension to the minimal singlet fermionic DM model whose entire parameter 
space (except a resonance region) excluded by the latest direct detection experiments.  
It is found out that there are viable regions in the parameter space which evade direct detection upper bounds
and respect the observed DM relic density by WMAP/Planck.
We also found that the Fermi-LAT upper limits on the DM annihilation cross section into $b\bar b$ 
can exclude small regions of the viable parameter space which elude direct detection experiments.
This model exemplifies a case within the WIMP paradigm whose DM candidate can escape direct detection 
experiments nontrivially.      
Such models are interesting to be studied in collider experiments like the LHC.

\end{abstract}
\vfill
\vfill
{\footnotesize\noindent }

\end{titlepage}



\section{Introduction}
\label{int}
One of the underlying questions in astro-particle physics is about the 
nature of dark matter (DM).
The current evidence for the existence of DM is ample and all 
of them are inferred from DM gravitational 
effects \cite{Bertone:2004pz,Bergstrom:2012fi}.
The standard model (SM) of particle physics has been remarkably successful in many aspects
but concerning the nature of DM, it is doomed to failure.
Although the present value of the DM relic density is measured with good precision 
by WMAP \cite{Hinshaw:2012aka} and Planck \cite{Ade:2013zuv}, we do not know yet 
the type and strength of the fundamental interactions of DM with the baryonic matter. 

So far, the current dedicated direct detection experiments have 
not manifested any sign of DM interaction with ordinary matter. 
However, the latest results by XENON1t \cite{Aprile:2017iyp} and 
LUX \cite{Akerib:2016vxi,Akerib:2017kat} for the DM-nucleon 
spin-independent/dependent cross section provide us with the stringent 
upper limits on the DM-nucleon cross sections for DM mass in the range of $\sim 10$ GeV up to $\sim$ a few TeV. 
The experimental upper bounds on the cross section can exclude or at least 
constrain various DM models which are attempting to explain the observed DM relic density.

One popular and noted paradigm for the DM is the weakly interacting massive
particles (WIMP) produced via freeze-out mechanism \cite{Lee:1977ua} where 
DM candidate has weak-scale interaction with the SM particles.
In typical WIMP models, new elementary or composite particles play the role of DM candidate.
Since the beginning of the construction of the WIMP models, the interaction of DM with 
normal matter through the SM Higgs portal has been an appealing possibility.   
The discovery of the Higgs boson \cite{Aad:2012tfa,Chatrchyan:2012xdj} reinforced 
the idea of the Higgs portal DM models. 
The Higgs discovery has been so welcome but its unnatural small mass 
uncovers the Higgs naturalness problem, for an interesting read in this 
regards one can consult \cite{Giudice:2017pzm}.   
One might then entertain the WIMP paradigm as a weak-scale solution to the Higgs 
naturalness problem as well, see for instance \cite{Kim:2018ecv,Kainulainen:2015raa,Ghorbani:2015xvz}.   

In building the WIMP models, the most minimal extensions to the SM look rationale to begin with. 
For a recent review on the various DM simplified models see \cite{Morgante:2018tiq} and references therein.
The singlet scalar DM model \cite{Silveira:1985rk,McDonald:1993ex,Burgess:2000yq} and 
the singlet fermionic DM model \cite{Kim:2008pp,Ghorbani:2014qpa,Ghorbani:2017qwf} 
exemplify the most minimal renormalizable WIMP models. 
The direct detection experiments have reached an unprecedented upper limits on the cross section,
such that we witness today that the most minimal DM models \cite{Burgess:2000yq,Kim:2008pp}
are getting excluded partially in the case of scalar DM \cite{Athron:2017kgt} or entirely 
in the case of fermionic DM \cite{Ettefaghi:2017vbh}.
In \cite{Ghorbani:2014gka} it is demonstrated that with two real scalar WIMPs 
coupled to the Higgs doublet, a viable DM mass above the Higgs resonance 
up to $\sim 200$ GeV is accessible. Some nonminimal extensions to 
the SM in order to save the singlet scalar DM are studied in \cite{Bhattacharya:2017fid,Casas:2017jjg}. 

In the present study, we extend the earlier work with a singlet fermionic DM \cite{Kim:2008pp} 
to a model with two singlet fermion WIMPs ($\psi_1$ and $\psi_2$), where the light one is stable and 
thus plays the role of the DM candidate. The heavy WIMP is sufficiently short-lived with no contribution to the DM relic density. 
A singlet scalar, $\phi$, acting as a mediator makes the DM-Higgs 
interaction possible. 
Due to a new interaction term $\sim g_{12} \bar \psi_1 \psi_2 +\text{(h.c)}$ 
in our model, the model possesses an interesting characteristic 
even in the limit of large WIMP mass splitting. It is found that adding one more Dirac 
fermion field can change the viable parameter space nontrivially. 
It is changed in such a way that the DM candidate can leave the detector without 
leaving any traces in the detector. 

The outline of the paper is as follows. In the next section, our DM model described. 
Lifetime of the heavy WIMP is evaluated numerically in section \ref{lifetime}. 
Numerical analysis on the DM relic density and the coannihilation effects are given 
in section \ref{Analysis}. The main results of this study obtained in section \ref{direct} 
by imposing direct detection bounds. We finish with a conclusion. In the Appendix, DM annihilation 
cross section formulas with two particles in the final state are provided.

\section{The model} 
\label{model}
In this section we introduce a model as a minimal extension to the SM which 
contains two gauge singlet Dirac fermions ($\psi_1,\psi_2$) one of which plays the role of 
the DM candidate, and a real singlet scalar ($\phi$) as a mediator with the Lagrangian, 

\begin{equation}
\label{DM-lag}
{\cal L}_{\text{DM}}   = \bar \psi_1 (i {\not}\partial-m_1)\psi_1 + \bar \psi_2 (i {\not}\partial-m_2)\psi_2 
       + g_{1}~ \phi \bar{\psi}_{1} \psi_{1} + g_{2}~ \phi \bar{\psi}_{2} \psi_{2}
                          +(g_{12}~ \phi \bar{\psi}_{1} \psi_{2} + \text{h.c.}) \,.
\end{equation}
The part of the Lagrangian which incorporates the new singlet scalar and the SM Higgs doublet is,
\begin{equation}
 {\cal L}(H,\phi )  =  \frac{1}{2} (\partial_{\mu} \phi)^2 - \frac{m^{2}}{2} \phi^2 
                       -\frac{\lambda_4}{24} \phi^4 - \frac{\lambda_1}{2} \phi H^{\dagger}H -\frac{\lambda_2}{2} \phi^2 H^{\dagger}H 
                       + \mu^{2}_{H} H^{\dagger}H - \frac{\lambda_H}{4} (H^{\dagger}H)^2 \,.
\end{equation}
As it is evident from the Lagrangians above, the only way for DM particle to interact 
with the SM particles is via a Higgs portal. Any other type of DM couplings to the SM particles 
are assumed to be negligible. 
It is well known that the Higgs doublet develops a non-zero vacuum 
expectation value ({\it vev}) as $v_0 = 246$ GeV, and we make an innocuous assumption 
where the singlet scalar acquires a zero {\it vev}.  
In the unitary gauge we write down the Higgs doublet 
as $H =  \left(0~~(v_0+ h^\prime)/\sqrt{2}\right)^{T}$. 
The mass matrix for the scalars is not diagonal due to 
the term linear in the $\phi$ field. The mass matrix at tree level reads

\begin{equation}\label{offmass}
 \mathbf{M^2} =
\begin{pmatrix}
 M^2_{h^\prime} & \frac{1}{2} M^2_{\phi h^\prime}\\
\frac{1}{2} M^2_{\phi h^\prime} &  M^2_{\phi} \\ 
\end{pmatrix}\,.
\end{equation}
where, 
\begin{equation}
 M^2_{h^\prime} = \frac{1}{2} \lambda_H v^2_0 \,,~~~  M^2_{\phi h^\prime} =  
 \lambda_1 v_0 \,,~~~ M^2_{\phi} = m^2 +\frac{1}{2} \lambda_2 v_0 \,. 
\end{equation}

To find the mass eigenstates we need to diagonalize the mass matrix and this can be achieved 
by rotating the relevant fields appropriately, 
\begin{equation}
h = \sin(\theta) ~\phi + \cos(\theta)~h^\prime\,, 
~~~~~~~~s = \cos(\theta)~\phi - \sin(\theta)~h^\prime\,, 
\end{equation}
such that the mixing angle $\theta$ satisfies the relation,
\begin{equation}
\tan{2 \theta} =  y\,,~~~~ y = \frac{2M^2_{\phi h}}{M^2_{h} - M^2_{\phi}} \,.
\end{equation}
The mass eigenvalues are obtained as the following,
\begin{equation}
m^{2}_{h,s} = \frac{M^{2}_{h^\prime}+M^{2}_{\phi}}{2}\pm \frac{M^{2}_{h^\prime}-M^{2}_{\phi}}{2} \sqrt{1+y^2}\,, 
\end{equation}
in which we choose $m_{h} = m_{H}$ for the SM Higgs mass with its measured value 125 GeV, and 
$m_{s}$ for the physical mass of the singlet scalar as a free parameter. 
It is possible to obtain two couplings in terms of the mixing angle and the physical masses of the scalars,
\begin{equation}
 \lambda_1 = \frac{\sin(2\theta)}{2 v_0} (m^2_{s} - m^2_{h}) 
 \,,~~~ \lambda_H = \frac{m^2_{s} \sin^2(\theta)+m^2_{h} \cos^2(\theta) }{v^2_0/2} \,.
\end{equation}
We assume that $m_1 < m_2$, therefore $\psi_1$ is the stable fermion and the DM candidate, and $\psi_2$ is its 
heavier partner, hence, we set $m_{\text{DM}} = m_1$.
The mass splitting of the fermions is defined as $\Delta = m_2 - m_1$.  
As a set of independent free parameters in the model 
we take $m_{s}, m_1, \Delta,\theta, g_1, g_2 , g_{12}, \lambda_2$, and $\lambda_4$.
Since our computations in this work are performed at tree level in perturbation theory, 
the coupling $\lambda_4$ does not come into play.
We are then left with eight free parameters.
The stability of the potential demands $\lambda_4 > 0$, $\lambda_H > 0$ and 
in case $\lambda_2 < 0 $ then we should have $\lambda_4 \lambda_H > 6 \lambda^2_2$.

\section{Lifetime of the heavy WIMP}
\label{lifetime}
The minimal model presents two fermionic WIMPs, where its heavy component 
can decay into the light partner (DM) and the SM Higgs as $\psi_2 \to \psi_1 h$. 
Depending on the size of the mass splitting the SM Higgs may decay successively as 
$h \to \bar f f, W^+ W^-, ZZ, ss$, where $f$ stands for the SM fermions. 
If the mass splitting is assumed to be $\Delta < m_{H}$, then the intermediate Higgs is off-shell 
and its kinematically allowed modes are decays into a pair of leptons or quarks (excluding the top quark). 

In case the lifetime of the heavy fermion is much less than the age of the Universe, then it has no contribution
to the present DM relic density. It is therefore necessary to see this condition is fulfilled 
over the viable parameter space. To this end, we give an expression for 
the decay width of the heavy fermion. Given the assigned 
momenta to the particles in the decay as $\psi_2(p_2) \to \psi_1(p_1) \bar f(p_3) f(p_4)$, it
reads
\ba
\Gamma = \int \int dt~du~ \frac{N_c \sin^2 2\theta~g^2_{12}~ m^2_f}{256\pi^3 m^3_2 v^2_h}
\Big[2m_1 m_2 (m^2_1+m^2_2)-(m^2_1+m^2_2)^2 - t m_1 m_2 + 2m^2_f (\Delta^2-t) 
\nonumber\\
+\frac{3}{2} t (m^2_1+m^2_2) - \frac{1}{2} t^2 \Big] \times
\Big|\frac{1}{(t-m^2_{s})+i\Gamma_{s} m_{s}}-\frac{1}{(t-m^2_{h})+i\Gamma_{h} m_{h}}\Big|^2\,,
\ea
where $N_c$ is the number of color for the SM fermion, and 
the Mandelstam variables are defined as $t=(p_3 + p_4)^2$ and $u=(p_2 + p_3)^2$. The decay rate 
of the heavy fermion is then $\tau = 1/\Gamma$. We compute the decay rate numerically 
utilizing the program {\tt CalcHEP} \cite{Belyaev:2012qa} for three distinct values of 
the mass splitting, $\Delta = 10, 50$ and $100$ GeV. The relevant input parameters are chosen as 
$g_{12} = 1.$, $\sin \theta = 0.05$ and $m_{s} = 300$ GeV.
\begin{figure}
\begin{center}
\includegraphics[width=.45\textwidth,angle =-90]{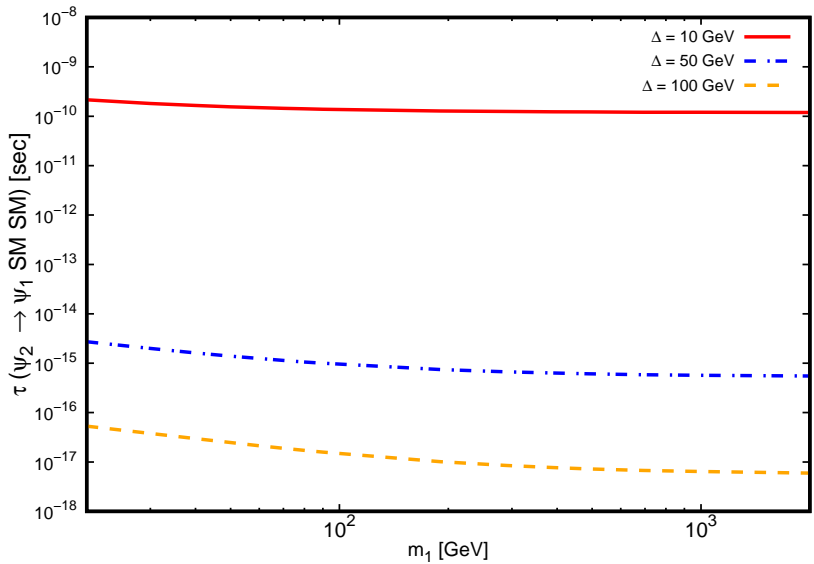}
\end{center}
\caption{The lifetime of the heavy fermion is provided for three values of 
the mass splitting, $\Delta = 10, 50, 100$ GeV. Chosen input parameters are: $g_{12} = 1$, $\sin \theta = 0.05$ and $m_s = 300$ GeV.} 
\label{lifetime-decay}
\end{figure}  
For the DM mass in the range 30 GeV $< m_1 <$ 2 TeV, it is apparent in Fig.~\ref{lifetime-decay}  
that the decay rate is much smaller than the age of the Universe and therefore has no contribution to 
the DM relic density.

\section{Thermal relic density and coannihilation effects}
\label{Analysis}
One natural way to explain the present DM relic density within the WIMP paradigm, is the 
production of thermal relic. In this process, WIMP particle(s) is in thermal equilibrium in the 
early Universe but, at a specific temperature named freeze-out temperature, 
it leaves the equilibrium and its density remains constant afterwards. 
The size of the relic density depends strongly on the freeze-out temperature, 
and the later depends in turn on the WIMPs annihilation cross sections in the early Universe. 

\begin{figure}
\begin{center}
\includegraphics[width=.6\textwidth,angle =0]{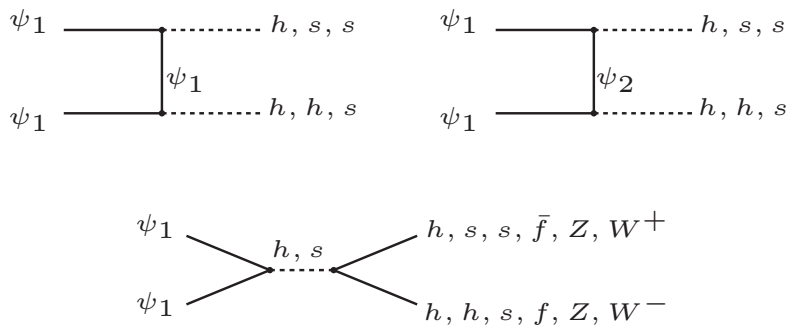}
\end{center}
\caption{Feynman diagrams for the DM annihilation processes are shown. Only diagrams 
with two particles in the final state are given here.} 
\label{Feynman-diag}
\end{figure}

In this work, we have a model with two participating WIMPs, in which the light one is our DM candidate. 
Besides the usual DM annihilation processes, there are events where DM candidate 
annihilates in tandem with another WIMP. This process is called coannihilation.
In Fig.~\ref{Feynman-diag} all possible annihilation diagrams for the DM candidate 
with two particles in the final state are displayed.
In the $s$-channel we have DM annihilation into $\bar f f,W^+W^-,ZZ,hh,hs,ss$ and 
in the $t$- and $u$-channel we have annihilation into $hh,hs,ss$. 
If one replaces a field $\psi_1$ with a field $\psi_2$ in the first two diagrams in Fig.~\ref{Feynman-diag}
then the coannihilation processes will be obtained.

In the case of two component WIMPs, to evaluate the DM relic abundance 
one needs to solve two coupled Boltzmann equations which
give us the time evolution of the WIMPs number densities.
In practice, it suffices to use a single Boltzmann equation with an effective (co)annihilation 
cross section instead of two Boltzmann equations \cite{Griest:1990kh,Edsjo:1997bg}. The resulting equation for the 
total number density, $n=n_1 + n_2$, is 
\begin{figure}
\begin{center}
\includegraphics[width=.45\textwidth,angle =-90]{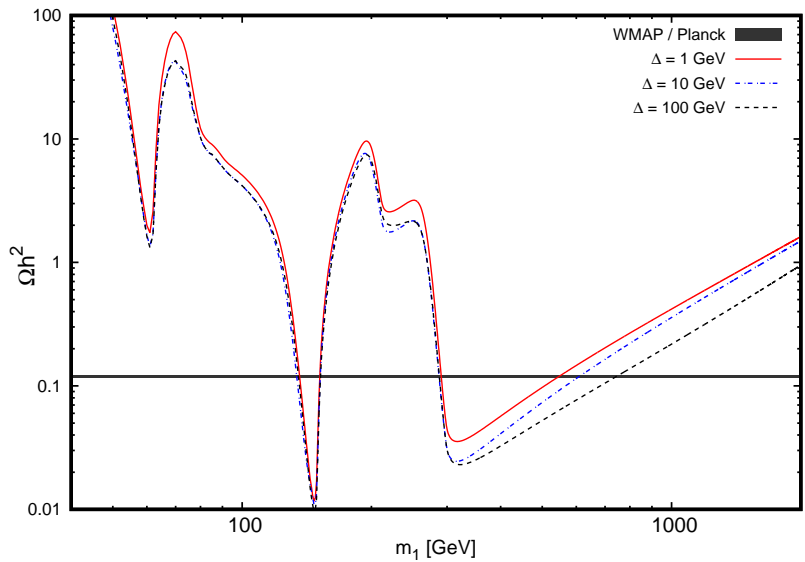}
\end{center}
\caption{DM relic density is shown as a function of DM mass for three values of 
the mass splitting, $\Delta  = 1, 10, 100$ GeV. Input parameters are: $g_1 = 0.9$, $g_2= 0.1$, $g_{12} = 0.9$,
$\lambda_2 = 0.1$, $\sin \theta = 0.05$ and $m_{s} = 300$ GeV.} 
\label{co-anni}
\end{figure}

\begin{equation}
\frac{dn}{dt}=-3Hn-\braket{\sigma_{\text{eff}}\,v}\left(n^2 - n^2_{eq} \right)\, ,
\end{equation}
in which $H$ is the Hubble constant and the effective cross section is given by
\begin{equation}
\sigma_{\text{eff}}=\frac{4}{g^2_{\text{eff}}}\left(\sigma_{11}+\sigma_{22}\left(1+\frac{\Delta}{m_{1}}\right)^{3}e^{-2\Delta/T}
+2\sigma_{12}\left(1+\frac{\Delta}{m_{1}}\right)^{3/2}e^{-\Delta/T} \right)\, ,
\end{equation}
where, $\sigma_{ij}$ denotes (co)annihilation 
cross section for the process, $\psi_i \psi_j \to \text{SM}$.  
The effective number of internal degrees of freedom is 
$g_{\text{eff}}=2+2\left(1+\frac{\Delta}{m_{1}}\right)^{3/2}e^{-\Delta/T}$.
In the Boltzmann equation, $\braket{\sigma_{\text{eff}}\,v}$ stands 
for the thermal averaged of the effective cross section times the relative DM velocity.
In the Appendix, formulas for $\sigma_{11}$ are given.

To obtain the DM relic density, one should solve the Boltzmann equation numerically. 
This is done in this work by exploiting the program {\tt MicrOMEGAs} \cite{Barducci:2016pcb}. 
In this stage we would like to study the dependency of the DM relic density on the mass 
splitting $\Delta$ in terms of the DM mass. 
In Fig.~\ref{co-anni} our results are shown for three values of the mass splitting, $\Delta = 1, 10, 100$ GeV.
The parameters used as input are, $g_1 = 0.9$, $g_2= 0.1$, $g_{12} = 0.9$,
$\lambda_2 = 0.1$, $\sin \theta = 0.05$ and $m_{s} = 300$ GeV.

It is evident from the plots in Fig.~\ref{co-anni} that the coannihilation effects which are 
relevant for small $\Delta$, increase the DM relic density slightly. The same behavior for the coannihilation 
effects is found within the supersymmetric models in \cite{Edsjo:2003us,Profumo:2006bx}. 
This can be compared with the case of scalar DM in the split-scalar model therein 
the relic density is decreased by the coannihilation effects \cite{Ghorbani:2014gka}. 
It is also seen that in the resonance region the coannihilation 
effects are essentially so small since in this region 
the DM annihilation cross section, $\sigma_{11}$, dominates 
the effective cross section. 

\section{Direct detection}
\label{direct}
Our main results concerning the direct detection bounds on the model parameter space 
are discussed in this section.
Direct detection experiments are set up in the hope that the strength of DM particle interaction 
with the nucleon is high enough such that finding DM footprints in the detector is feasible.
Recent results from XENON1t and LUX experiments suggest that we are not able to 
find a DM signal in the detector yet. However, the most stringent exclusion limits 
are granted by these experiments. 

Let us now find regions in our model parameter space where 
the DM candidate can evade the exclusion bounds. 
In order to find the DM-nucleon elastic scattering cross section, first one needs to have the 
DM interaction with ordinary matter in the quark level. The DM-quark interaction is possible 
in the present model by exchanging a SM Higgs or a singlet scalar via $t$-channel processes. 
In the limit of low momentum transfer, the relevant interaction can be described by an effective 
Lagrangian, 
\ba
{\cal L_{\text{eff}}} = c_{\text{q}}~\bar \psi_1  \psi_1~ \bar q q \,,  
\ea 
with the effective coupling, 
\ba
c_{\text{q}} = (g_1 \sin 2\theta) \frac{m_{\text{q}}}{2 v_{0}} (\frac{1}{m^{2}_{h}}-\frac{1}{m^{2}_{s}}).
\ea
To promote the effective Lagrangian to the nucleon level, we appeal to the approximation 
at the very small momentum transfer, where a quark level matrix element can be replaced by 
a nucleonic level matrix element introducing a proportionality form factor \cite{Ellis:2008hf,Crivellin:2013ipa}.
Therefore, we can ultimately obtain the spin-independent (SI) 
cross section for the DM-nucleon as 
\begin{equation}
 \sigma^{\text{N}}_{\text{SI}} = \frac{4\mu^2_{N} c^2_{N}}{\pi} \,,
\label{DDCS}
\end{equation}
where the reduced mass of the DM-nucleon is denoted by $\mu_{N}$ and, the coupling $c_{N}$ is given
in terms of the quark effective coupling $c_{q}$ and scalar form factors as 
\ba
c_{N} = m_{N} \sum_{q = u,d,s} f^{N}_{Tq} \frac{c_{q}}{m_{q}} 
+ \frac{2}{27} f^{N}_{Tg} \sum_{q = c,b,t}   \frac{c_{q}}{m_{q}} \,.
\label{CC}
\ea
\begin{figure}
\hspace{-1.5cm}
\begin{minipage}{0.4\textwidth}
\includegraphics[width=\textwidth,angle =-90]{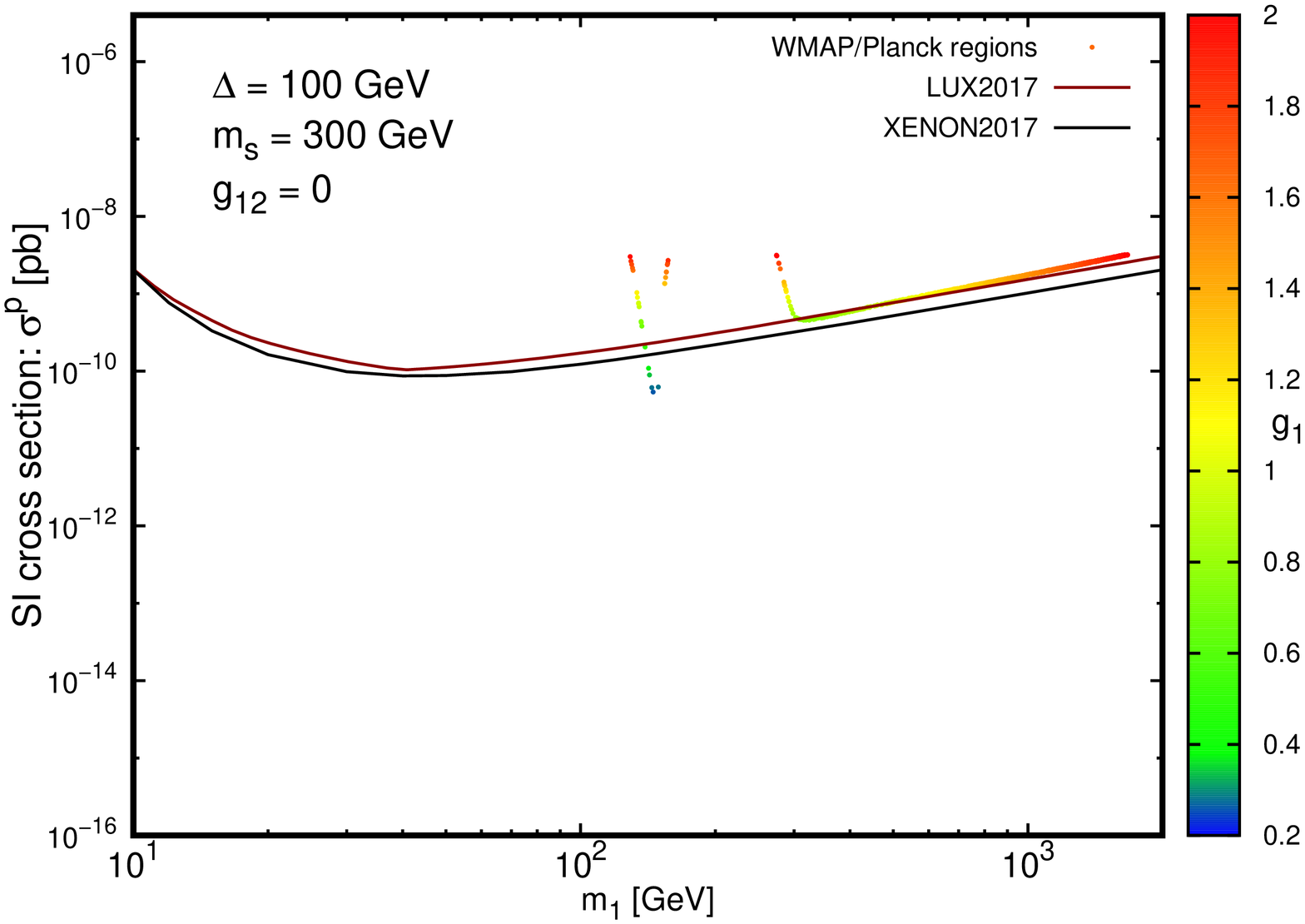}
\end{minipage}
\hspace{2.5cm}
\begin{minipage}{0.4\textwidth}
\includegraphics[width=\textwidth,angle =-90]{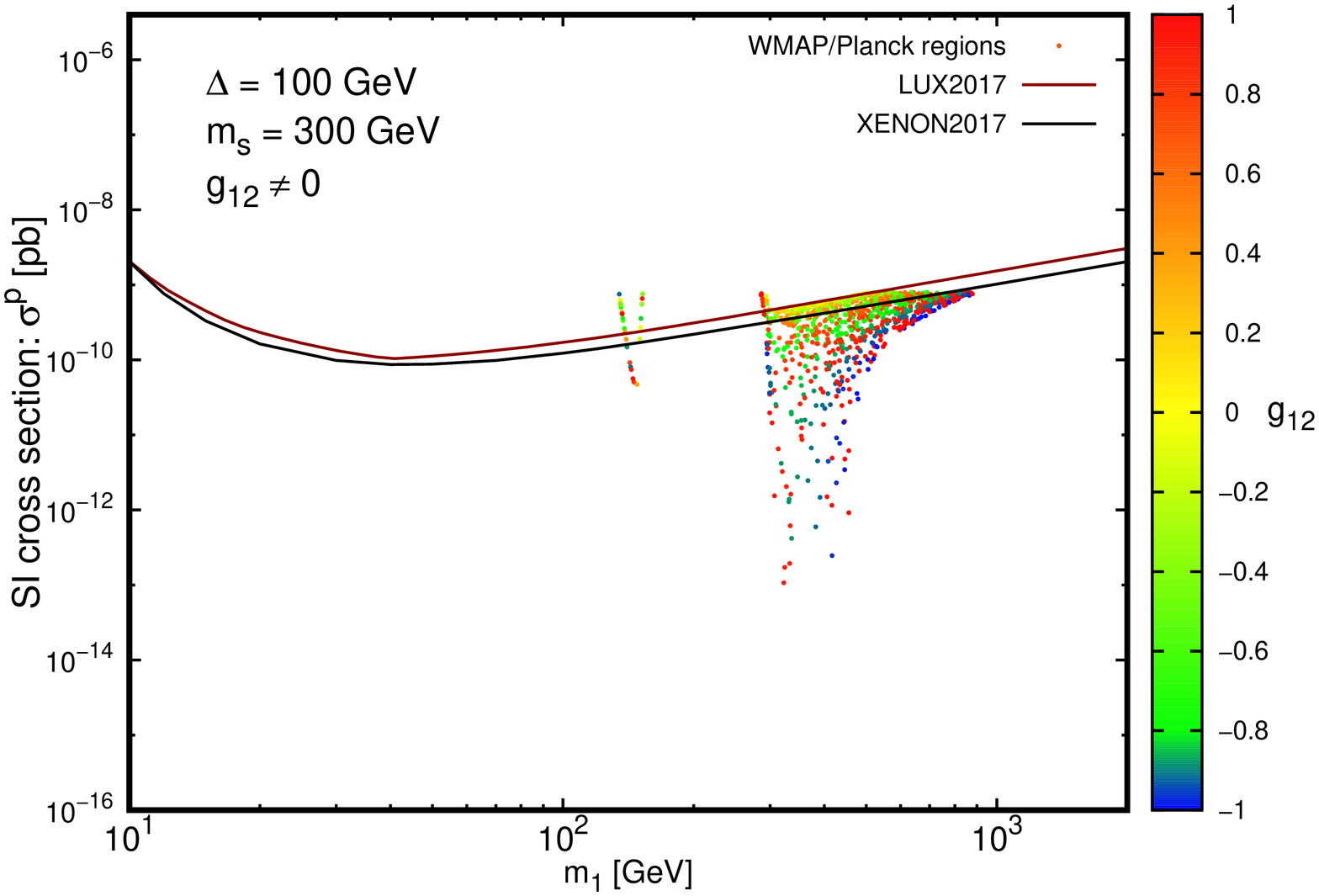}
\end{minipage}
\caption{The DM-proton (SI) scattering cross section as a function of DM mass for the mass splitting
$\Delta = 100$ GeV and singlet scalar mass $m_s = 300$ GeV. 
In the left panel $g_{12} = 0 $ and in the right panel $g_{12} \ne 0$. The mixing angle is fixed at $\sin \theta = 0.05$.}
\label{dirdel100}
\end{figure}
In the analysis to come, for the DM-proton elastic cross section we have used the scalar form factors below 
\ba
f^{p}_{u} = 0.0153,~~~~~~~ f^{p}_{d} = 0.0191, ~~~~~~~ f^{p}_{s} = 0.0447 \,.
\ea  

In the following numerical computations we always choose for the mixing angle $\sin \theta = 0.05$, and for 
the singlet scalar we select two distinct masses, $m_s = 80, 300$ GeV.

To begin, let us take a large value for the mass splitting, $\Delta = 100$ GeV, 
where according to the standard lore we expect the 
coannihilation effects to be relevant up to the DM masses such that $\frac{\Delta}{m_{\text{DM}}}\lesssim 10\%$. 
At first sight it seems that
the present model should reduce to the singlet fermionic model at the limit of large mass splitting. 

To check this out, we consider two cases (a) $g_{12} = 0$ and (b) $g_{12} \ne 0$,
and then move on to compute the DM-proton elastic 
scattering cross section as a function of the DM mass in the range $30$ GeV $< m_1 < 2$ TeV.
We set, for the first case $0 < g_1, g_2 < 2$, and for the second case  $-1 < g_1, g_2 , g_{12} < 1$.

In Fig.~\ref{dirdel100} direct detection cross sections are shown for the two cases
where only points consistent with the observed relic density are picked.
In the case with $g_{12} = 0$ the entire parameter space is excluded by direct detection bounds except
the resonance region around $m_1 \sim m_{s}/2$. This is in agreement with the results in \cite{Ettefaghi:2017vbh}. 
However, in case $g_{12} \ne 0$, besides the resonance region there are points in the parameter space
in the range 300 GeV $\lesssim  m_1 \lesssim 600$ GeV residing well below the exclusion bounds.
Since the coannihilation effects are believed to be tiny for $\Delta = 100$ GeV in the above 
mentioned mass range, a pertinent question to pose is that why in case $g_{12} \ne 0$ 
there are points in the parameter space which can evade exclusion bounds. To address this question 
we should first note that at the limit of large mass splitting the effective (co)annihilation 
cross section becomes $\sigma_\text{eff} \sim \sigma_{11}$.
Therefore, according to the annihilation cross section formulas in the Appendix, in case $g_{12} = 0$
then we have $\sigma_\text{eff} \propto g^2_{1}$, and when $g_{12} \ne 0$ then
the effective cross section depends on the coupling $g_1,g_{12}$ in various ways, $\sigma_\text{eff} \propto g^4_{1}, ...,g^4_{12}$.
On the other hand, the (SI) elastic scattering cross 
section, $\sigma^{\text{N}}_{\text{SI}}$, is proportional to $g^2_{1}$ as
seen from the formulas in Eq.~(\ref{DDCS}) and Eq.~(\ref{CC}).
\begin{figure}
\hspace{-1.5cm}
\begin{minipage}{0.4\textwidth}
\includegraphics[width=\textwidth,angle =-90]{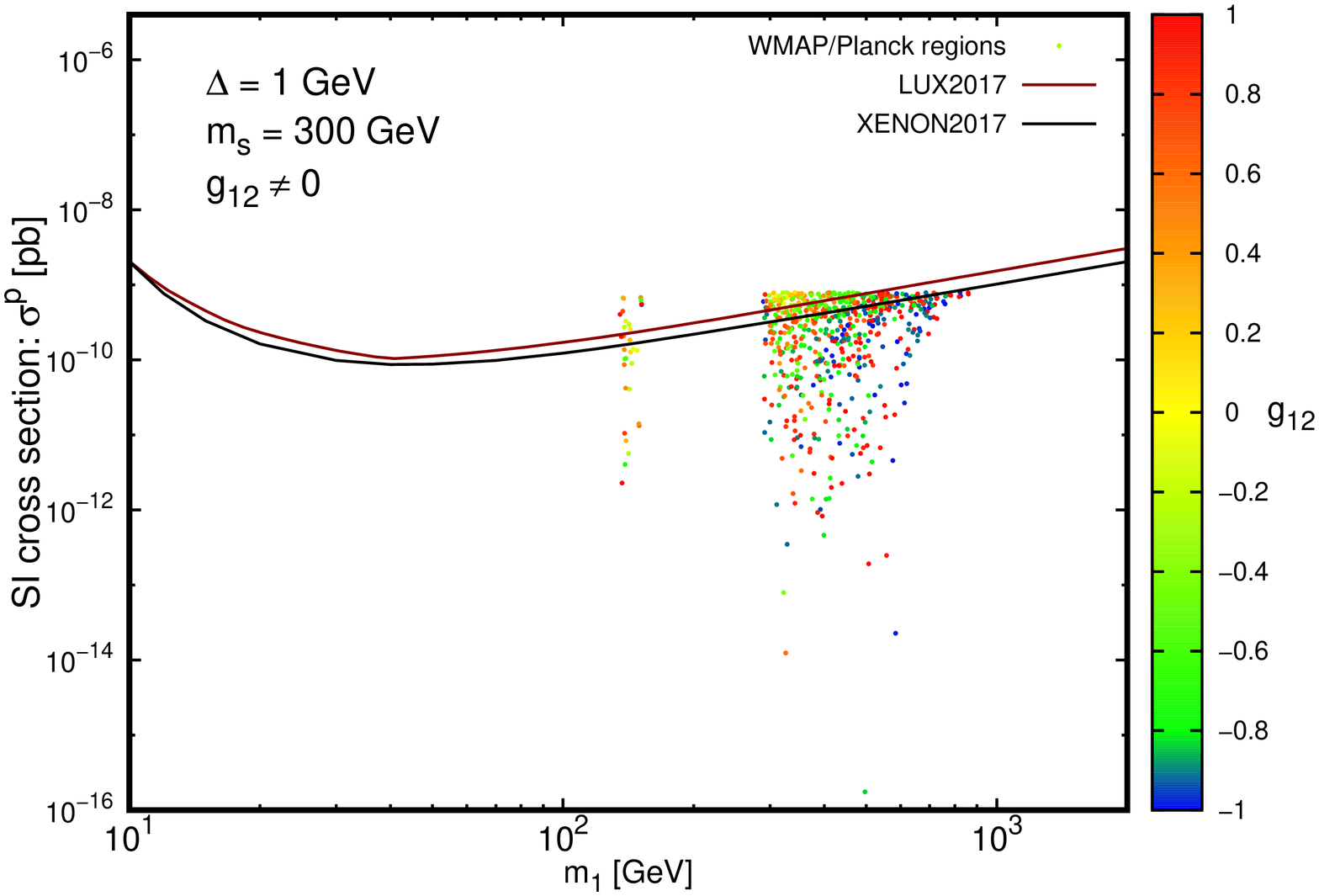}
\end{minipage}
\hspace{2.5cm}
\begin{minipage}{0.4\textwidth}
\includegraphics[width=\textwidth,angle =-90]{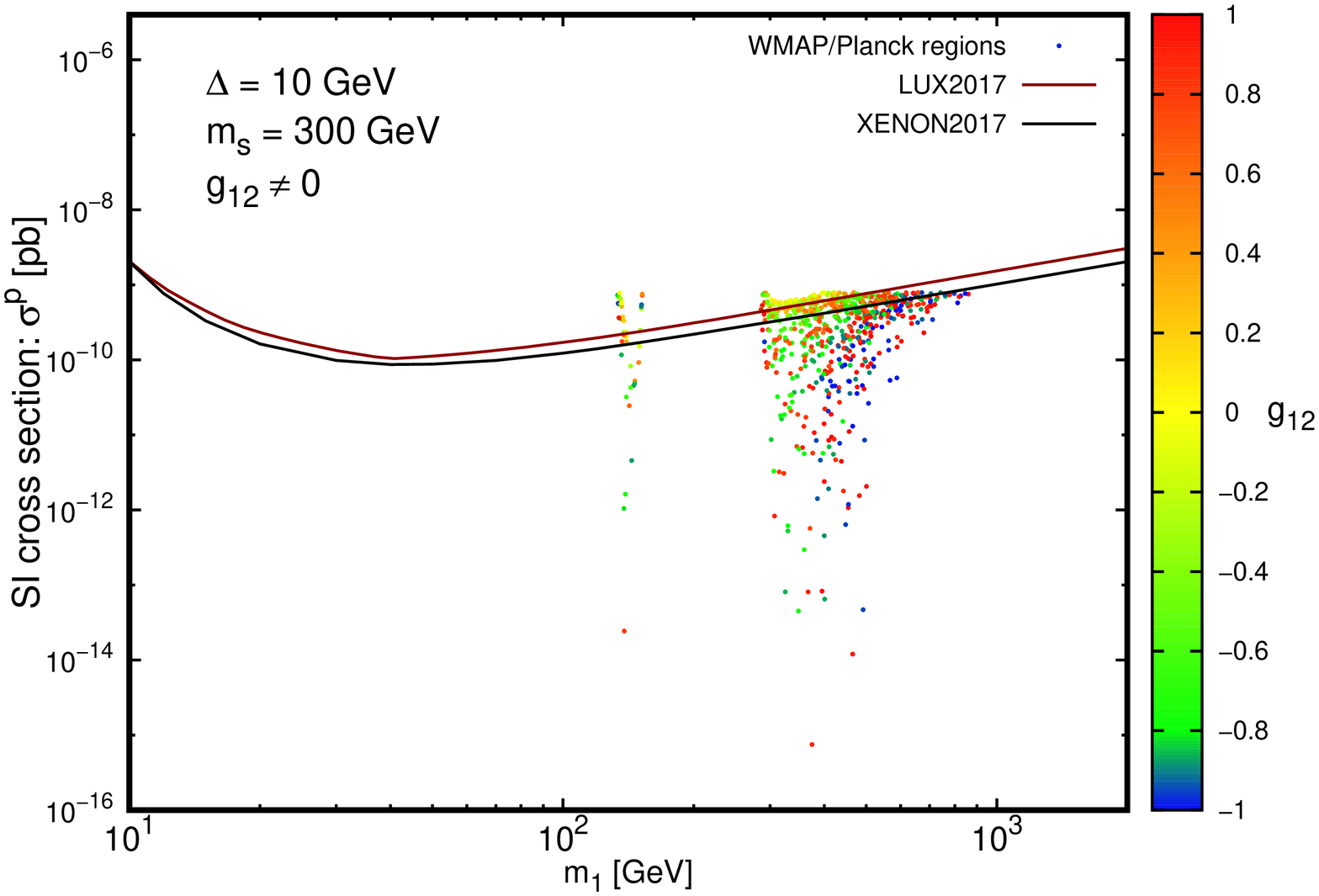}
\end{minipage}
\caption{The DM-proton (SI) scattering cross section as a function of DM mass for the mass splitting
$\Delta = 1, 10$ GeV. The singlet scalar mass is $m_s = 300$ GeV and the mixing angle is $\sin \theta = 0.05$.}
\label{dirdel1}
\end{figure}

It is not difficult now to explain our results: When $g_{12} = 0$, in order to get a large enough annihilation
cross section (to induce the observed relic density) large value for the coupling $g_{1}$
is required. This in turn gives rise to a large (SI) elastic scattering cross section 
which is also proportional to $g^2_{1}$ and hence, the entire parameter space get excluded.
When $g_{12} \ne 0$ it becomes possible to have a small
(SI) elastic scattering cross section (small $g_{1}$) and at the same time large
enough annihilation cross section, suitable to give the correct relic density. This is due to terms
proportional to powers of the coupling $g_{12}$, necessitated to enhance the annihilation cross section. 

Another important observation from Fig.~\ref{dirdel100} indicates that the viable DM mass
when $m_s = 300$ GeV, starts from $m_{\text{DM}} \sim 300$ up to higher masses for $g_{12} \ne 0$. 
The reason hinges on the fact that for the DM masses larger than 300 GeV, a new channel 
$\psi_1 \bar \psi_1 \to s s$ opens up for the DM annihilation which has dominant effects 
over the other already opened channels, namely, $\psi_1 \bar \psi_1 \to h h$ 
and $\psi_1 \bar \psi_1 \to s h$.
In fact, when the process $\psi_1 \bar \psi_1 \to s s$ opens up, we get contributing terms proportional 
to $g^2_1 g^2_{12} \cos^4 \theta$ or $g^4_{12} \cos^4 \theta$ while 
in the processes $\psi_1 \bar \psi_1 \to h h$ and $\psi_1 \bar \psi_1 \to h s$, the contributing 
terms to the new effect are proportional to $g^2_1 g^2_{12} \sin^4 \theta$ or $g^4_{12} \sin^4 \theta$, and
$g^2_1 g^2_{12} \cos^2 \theta \sin^2 \theta$ or $g^4_{12} \cos^2 \theta \sin^2 \theta$, respectively.
Since $\sin \theta = 0.05$, it is easy to see that the process $\psi_1 \bar \psi_1 \to s s$ 
brings in the dominant effects and the DM mass threshold at about 300 GeV becomes explicable. 

For smaller mass splitting, $\Delta = 1, 10$ GeV, the results are shown in Fig.~\ref{dirdel1}.
The results indicate almost the same characteristic as the case with $\Delta = 100$ GeV.   
\begin{figure}
\hspace{-1.5cm}
\begin{minipage}{0.4\textwidth}
\includegraphics[width=\textwidth,angle =-90]{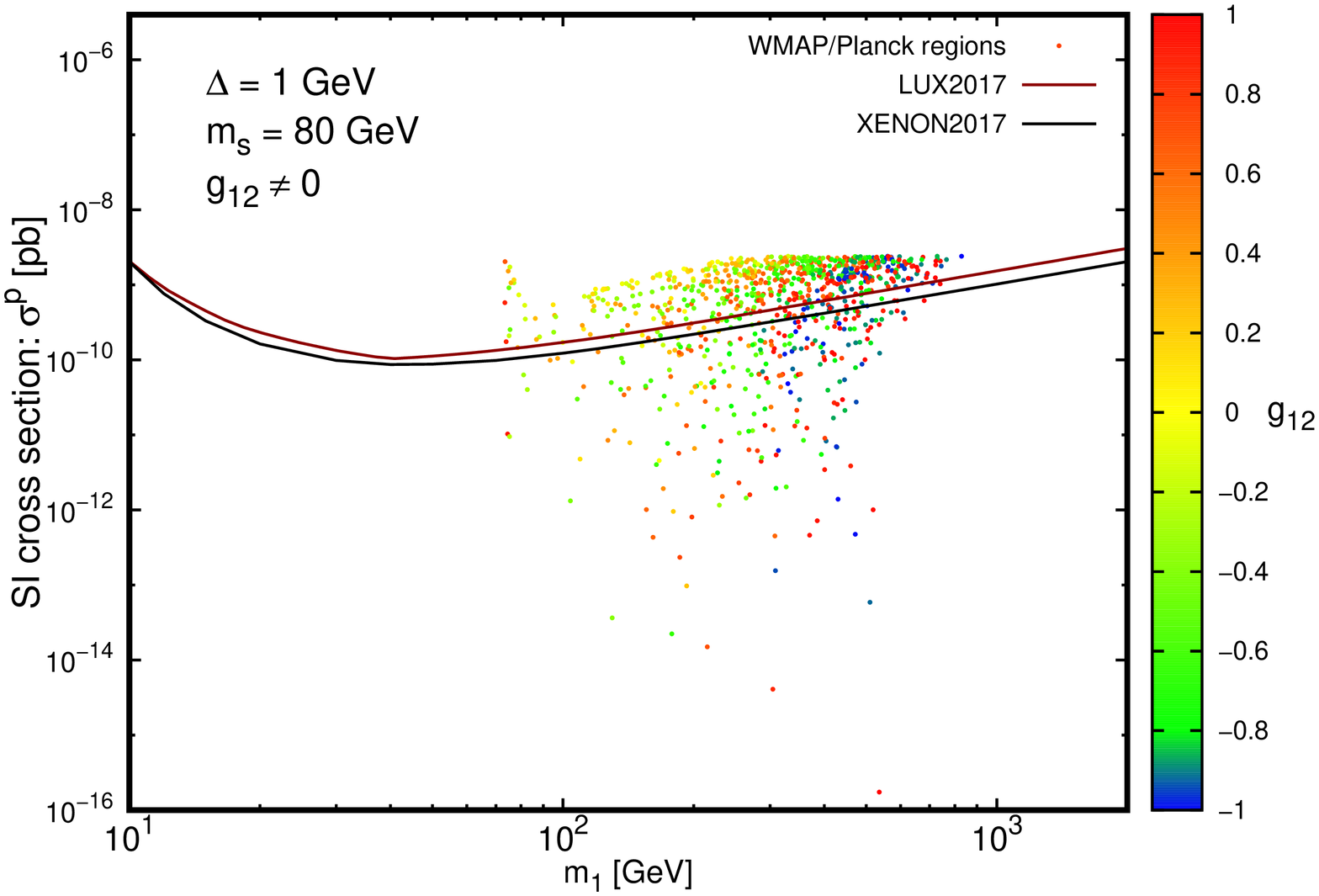}
\end{minipage}
\hspace{2.5cm}
\begin{minipage}{0.4\textwidth}
\includegraphics[width=\textwidth,angle =-90]{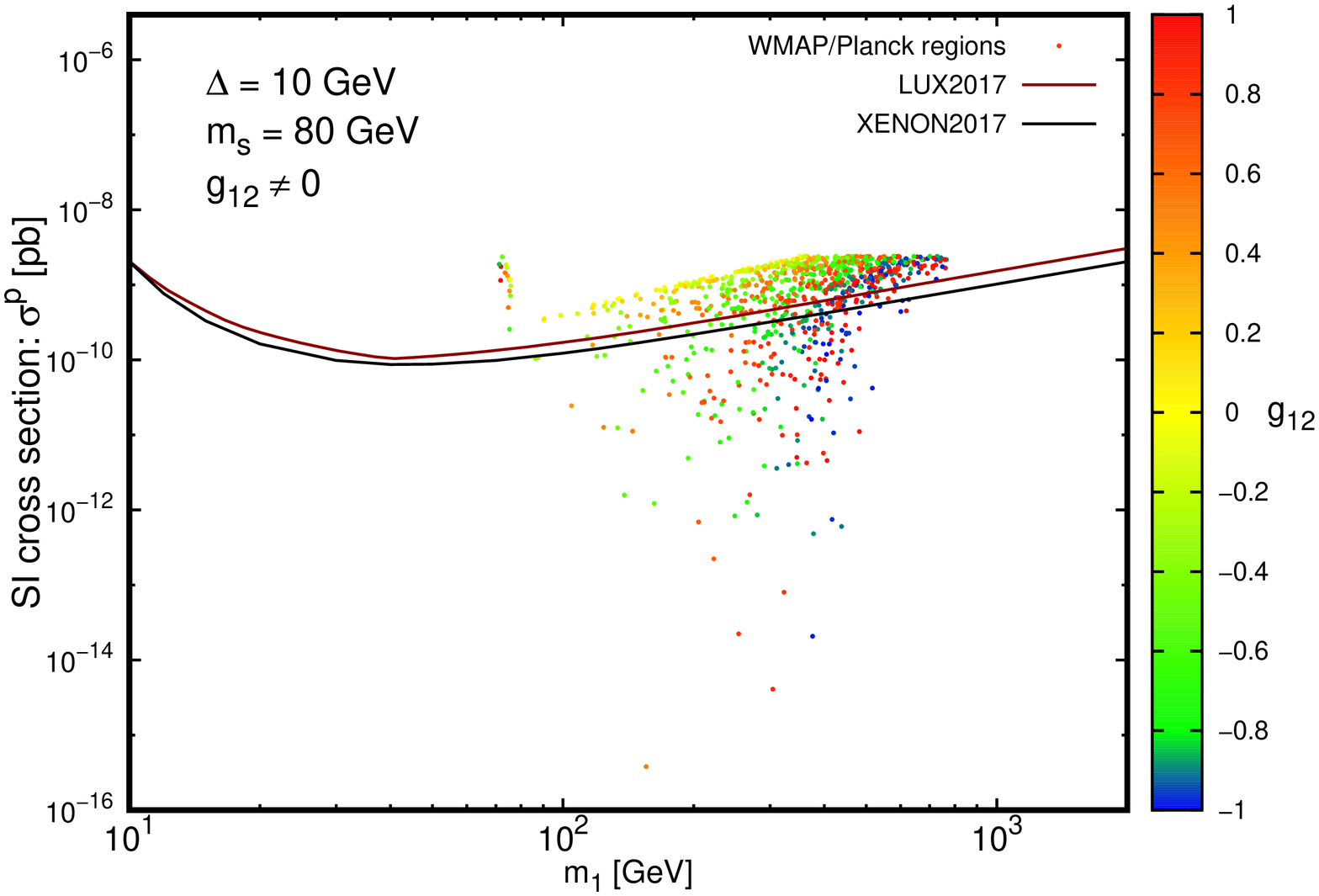}
\end{minipage}
\caption{The DM-proton (SI) scattering cross section as a function of DM mass for the mass splitting
$\Delta = 1, 10$ GeV, and the singlet scalar mass $m_s = 80$ GeV.}
\label{dirde-ms80}
\end{figure}

\begin{figure}
\begin{center}
\includegraphics[width=.45\textwidth,angle =-90]{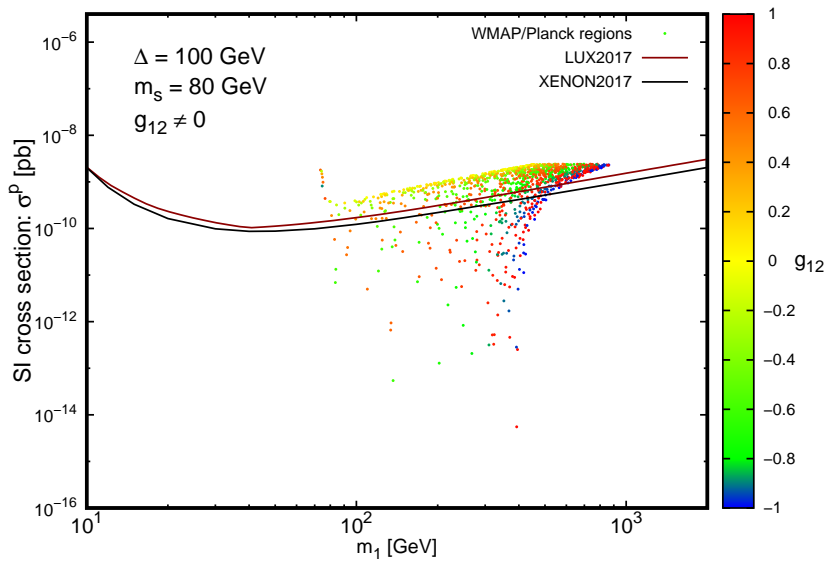}
\end{center}
\caption{The DM-proton (SI) scattering cross section as a function of DM mass for the mass splitting
$\Delta = 100$ GeV, and the singlet scalar mass $m_s = 80$ GeV.} 
\label{dirde-ms80-100}
\end{figure}

In the next step, we redo our computations taking a smaller mass for 
the singlet scalar, i.e., $m_s = 80$ GeV and the same values for the other input parameters. 
Our findings are given in Fig.~\ref{dirde-ms80} for $\Delta = 1, 10$ GeV, and 
in Fig.~\ref{dirde-ms80-100} for $\Delta = 100$ GeV.
The range of the viable DM mass is now broaden significantly, from $m_1 \sim m_s/2$ up to 
masses $\sim 600$ GeV in the case with $\Delta = 1$ GeV. At larger mass splitting 
that the coannihilation effects are less important the lower viable DM mass is pushed a little upward.  
Again, the same line of reasoning can be applied to explain the viable DM mass range. 

\begin{figure}
\hspace{-1.5cm}
\begin{minipage}{0.4\textwidth}
\includegraphics[width=\textwidth,angle =-90]{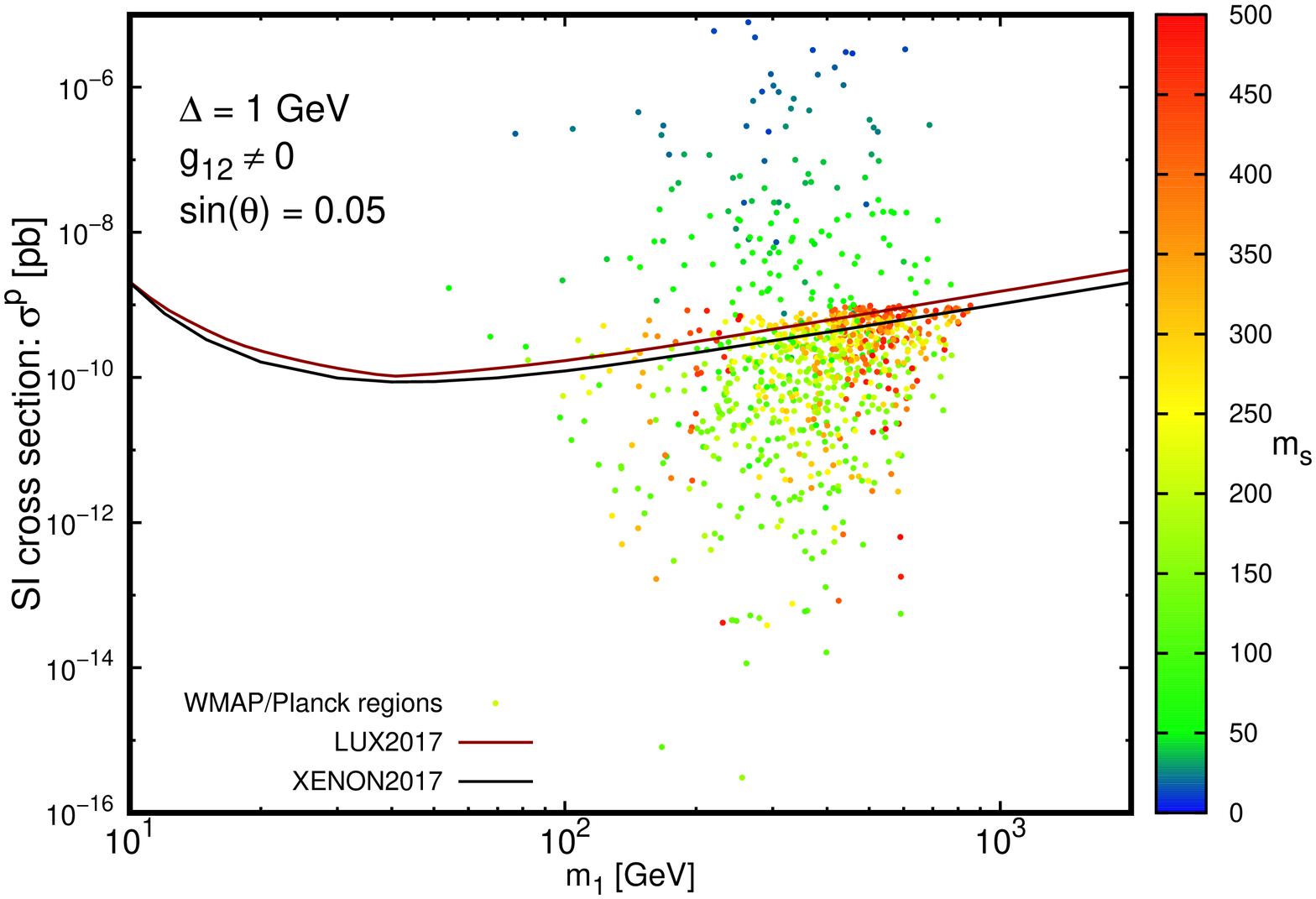}
\end{minipage}
\hspace{2.5cm}
\begin{minipage}{0.4\textwidth}
\includegraphics[width=\textwidth,angle =-90]{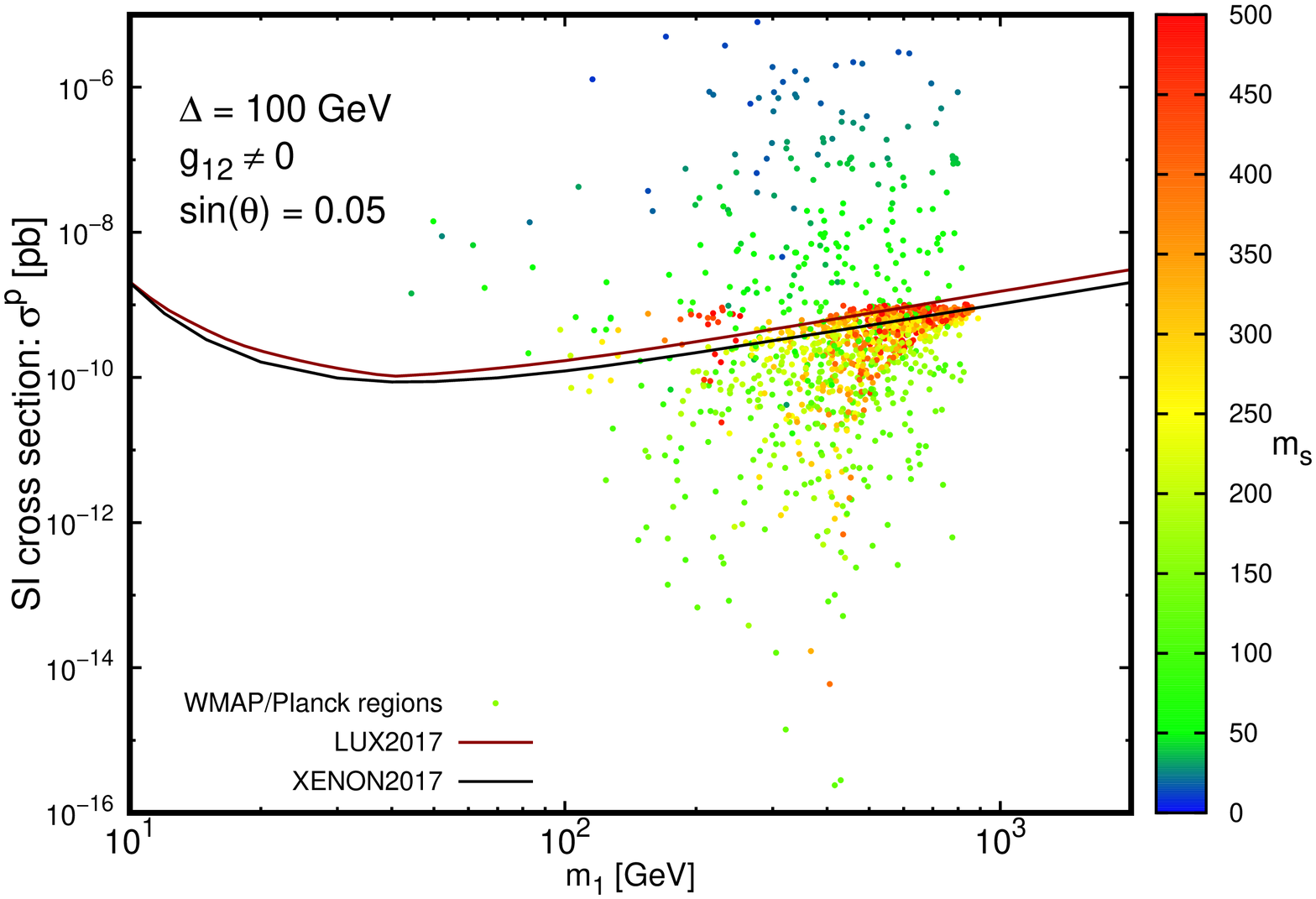}
\end{minipage}
\caption{The DM-proton (SI) scattering cross section as a function of DM mass for the mass splitting
$\Delta = 1$ GeV {\it (left)} and $\Delta = 100$ GeV {\it (right)}. The singlet scalar mass is 10 GeV $< m_s < 500$ GeV
and the mixing angle is $\sin \theta = 0.05$. The couplings are chosen as $-1 < g_1, g_2, g_{12} < 1$.}
\label{ms-sin05}
\end{figure}

\begin{figure}
\hspace{-1.5cm}
\begin{minipage}{0.4\textwidth}
\includegraphics[width=\textwidth,angle =-90]{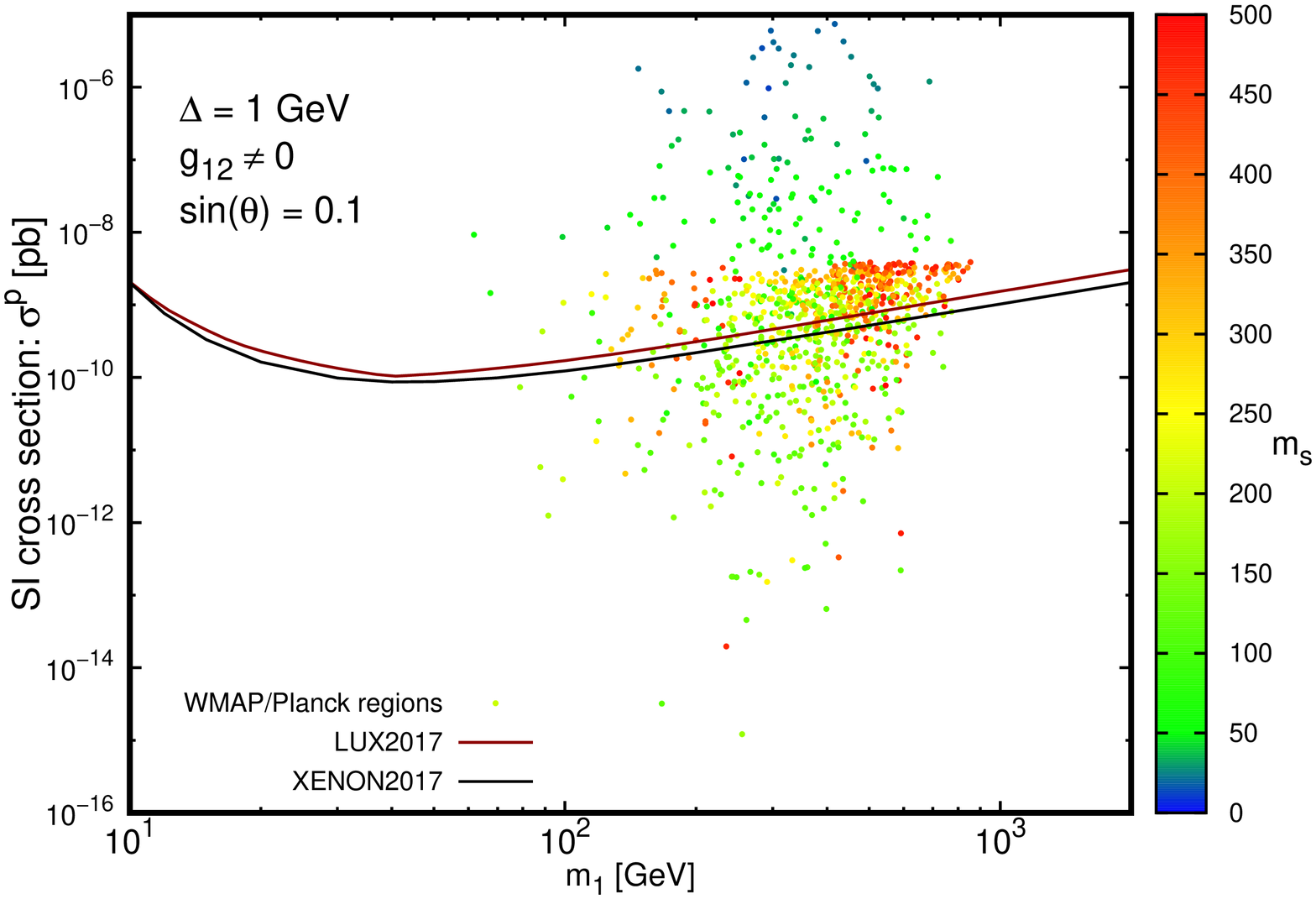}
\end{minipage}
\hspace{2.5cm}
\begin{minipage}{0.4\textwidth}
\includegraphics[width=\textwidth,angle =-90]{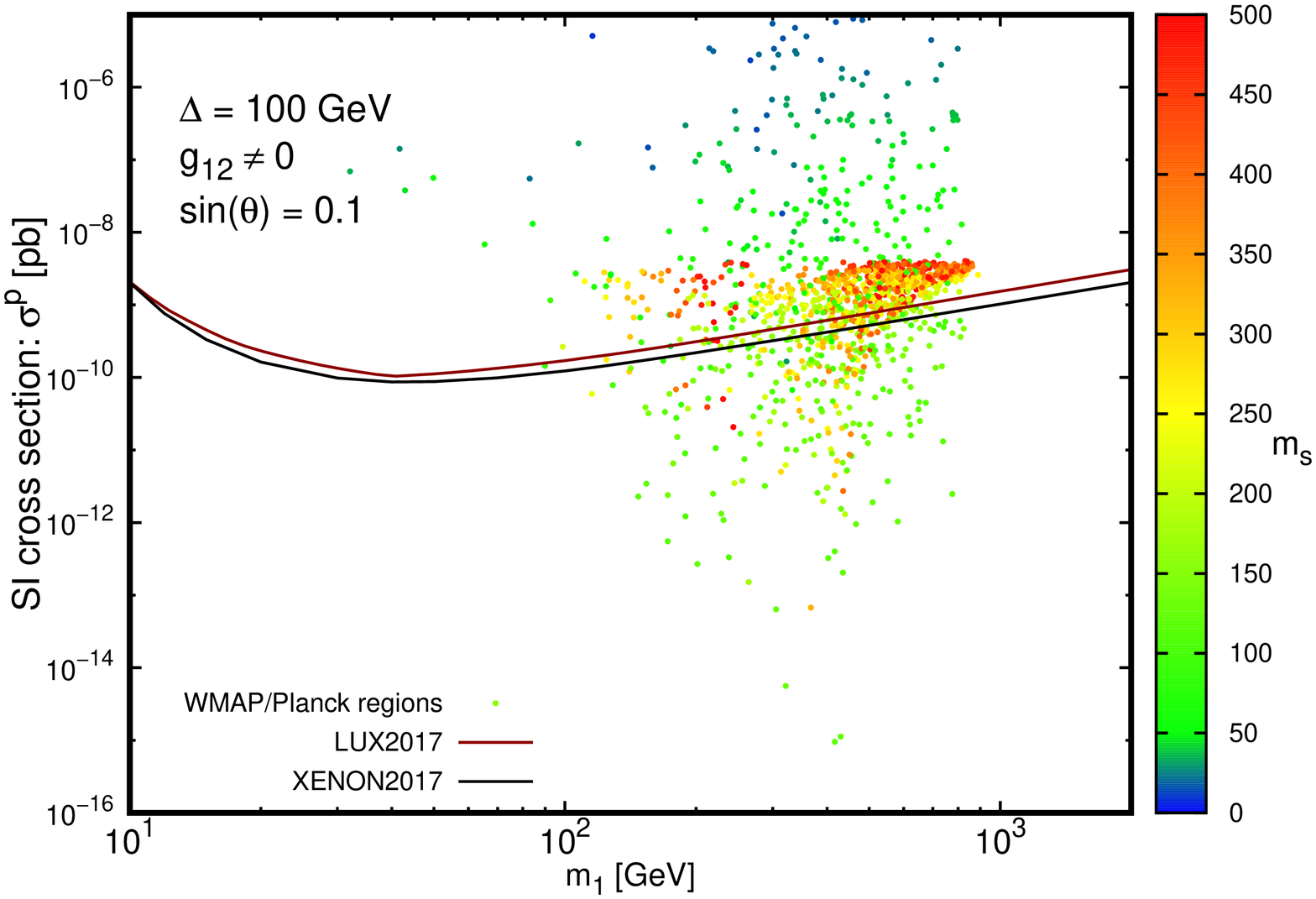}
\end{minipage}
\caption{The DM-proton (SI) scattering cross section as a function of DM mass for the mass splitting
$\Delta = 1$ GeV {\it (left)} and $\Delta = 100$ GeV {\it (right)}. The singlet scalar mass is 10 GeV $ < m_s < 500$ GeV
and the mixing angle is $\sin \theta = 0.1$. The couplings are chosen as $-1 < g_1, g_2, g_{12} < 1$.}
\label{ms-sin1}
\end{figure}

Finally, we take the singlet scalar mass in the range 10 GeV $< m_s <$ 500 GeV 
and compute the DM-nucleon scattering cross section for two distinct values of the mass splitting 
$\Delta = 1, 100$ GeV and for two values of the mixing angle, $\sin \theta = 0.05, 0.1$. 
The regions of the parameter space respecting the observed relic density are shown in Fig.~\ref{ms-sin05} with 
$\sin \theta = 0.05$ and in Fig.~\ref{ms-sin1} with $\sin \theta = 0.1$.  
The results indicate that singlet scalar masses with $m_s \lesssim 30$ GeV are excluded 
by direct detection upper bounds irrespective of our choices for the mass splitting and the mixing angle.

\section{Indirect Detection} 

The existence of DM in regions with high density like the Galactic Center and dwarf 
spheroidal satellite galaxies (dSphs) of the Milky Way might lead to self-annihilation of DM into energetic SM particles. 
This motivates indirect search for DM. 
{\it Fermi} Large Area Telescope (Fermi-LAT) collection of gamma-ray data for six years from Milky Way dSphs \cite{Ackermann:2015zua}
and the H.E.S.S. (High Energy Stereoscopic System) ground-based cherenkov telescopes 
with ten years of Galactic Center gamma-ray data collection \cite{Abdallah:2016ygi} 
do not indicate any significant gamma-ray excess. 

However, H.E.S.S., assuming Einasto and NFW DM density
profiles at the Galactic Center provides us with the upper bounds on the velocity-weighted annihilation 
cross section $\braket{\sigma v}$ for various channels, namely, annihilation into quark pair ($b\bar b,t\bar t$), gauge boson pair ($W^+W^-$)
and lepton pair ($\mu^+\mu^-,\tau \bar \tau$) channels. The strongest upper limit is obtained for a particle DM mass 
of about 1 TeV in the $\tau^+ \tau^-$ channel at $\sim 2\times10^{-26}$ $\text{cm}^3\text{s}^{-1}$ \cite{Abdallah:2016ygi}. 
In addition, Fermi-LAT gives us the upper limits on the annihilation cross sections for the same channels including
$e^+ e^-$ channel \cite{Ackermann:2015zua}. 

For the regions of the parameter space we are interested in the Fermi-LAT upper limits are slightly stronger, thus these limits 
are applied in our probe over the viable regions. 

\begin{figure}
\hspace{-1.5cm}
\begin{minipage}{0.4\textwidth}
\includegraphics[width=\textwidth,angle =-90]{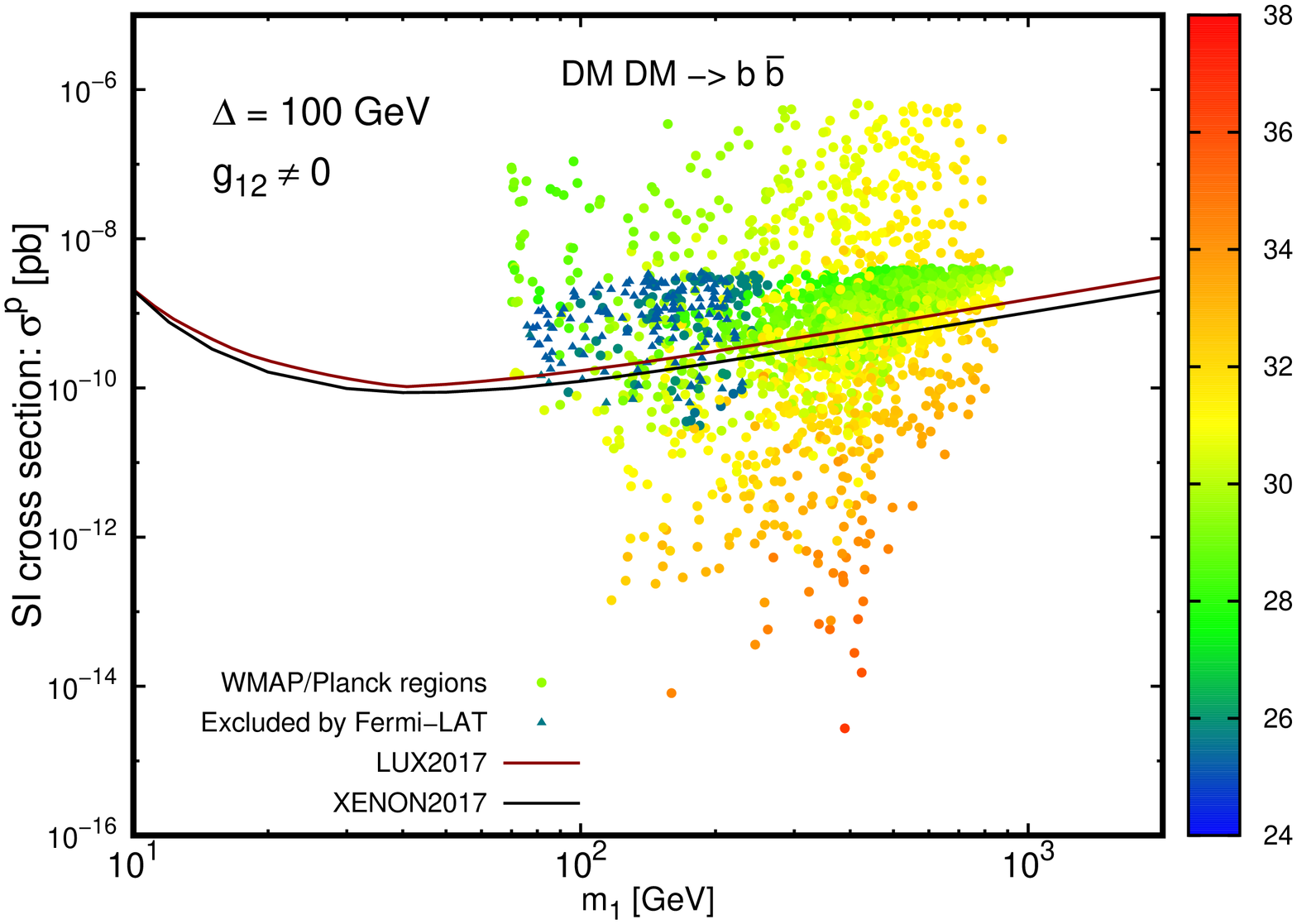}
\end{minipage}
\hspace{2.5cm}
\begin{minipage}{0.4\textwidth}
\includegraphics[width=\textwidth,angle =-90]{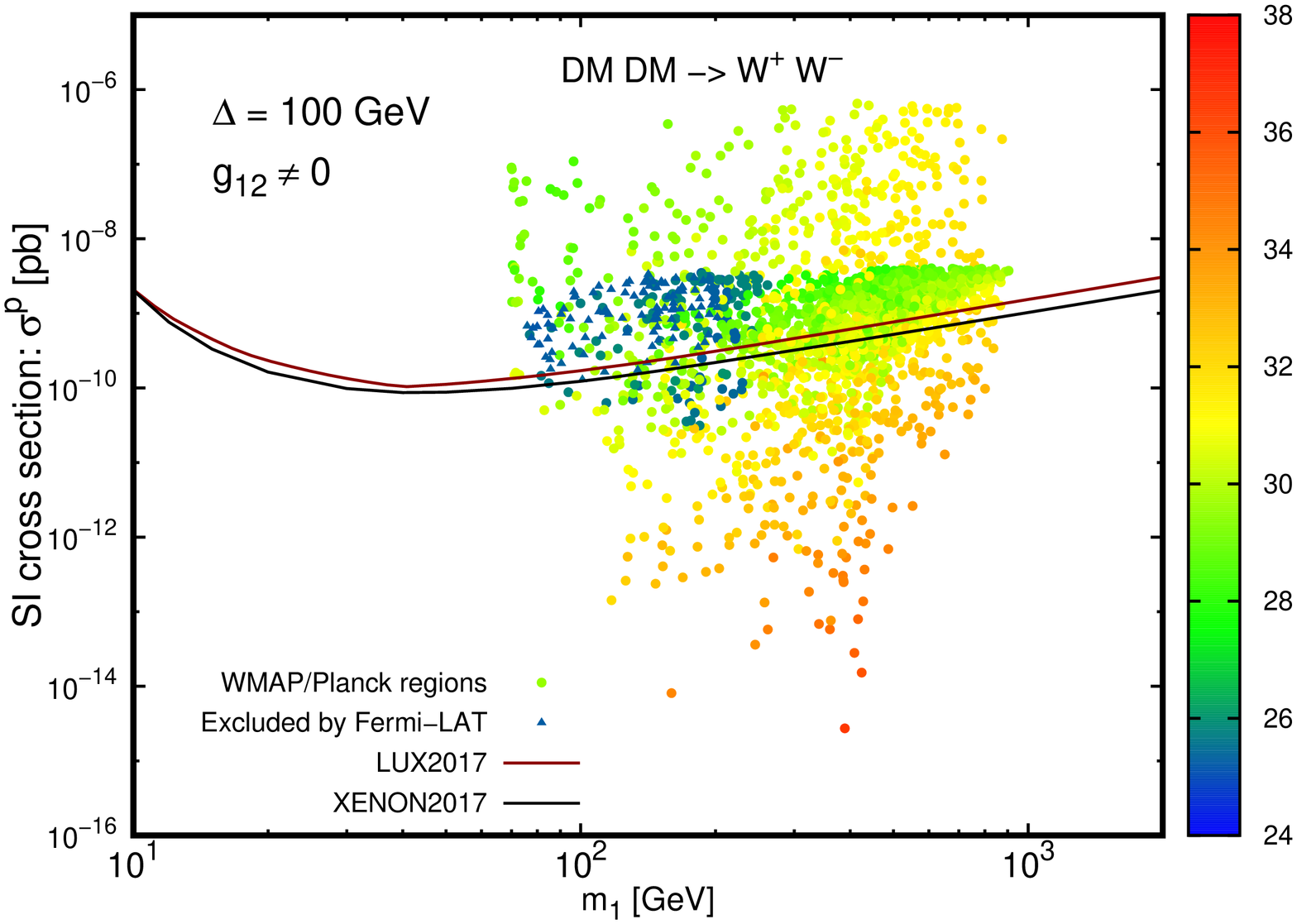}
\end{minipage}
\caption{The DM-proton (SI) scattering cross section as a function of DM mass for the mass splitting
$\Delta = 100$ GeV. The vertical color spectrum indicates the quantity $-log_{10} \braket{\sigma v}$ for 
DM annihilation into $b\bar b$ {\it (left)} and DM annihilation into $W^+W^-$ {\it (right)}. }
\label{indirect-bbww}
\end{figure}

In our numerical computations we pick the free parameters as $\Delta = 100$ GeV, $\sin \theta = 0.1$, 
30 GeV $< m_s <$ 500 GeV, 10 GeV $< m_1 < 2000$ GeV and $-1 < g_i < 1$. 
In Fig.~\ref{indirect-bbww} our results are shown for the DM-proton cross section in terms of the DM mass, and also 
the quantity $-log_{10} \braket{\sigma v}$ is given for the DM annihilation into $b \bar b$ and $W^+ W^-$ constrained by the 
Fermi-LAT upper limits.
It is found that the Fermi-LAT constraints on $\braket{\sigma v}_{b\bar b}$ is slightly 
stronger than that on $\braket{\sigma v}_{W^+ W^-}$, such that a small portion of the viable parameter space gets excluded 
by the Fermi-LAT upper limits on $\braket{\sigma v}_{b\bar b}$ but Fermi-LAT upper bounds on $\braket{\sigma v}_{W^+ W^-}$ can 
only exclude regions in the parameter space which is already excluded by direct detection experiments.
We also checked and found that other channels probed by Fermi-LAT i.e., $e^+e^-, \mu^+ \mu^-$ and $\tau^+ \tau^-$,
are not strong enough to exclude any region of interest in our model parameter space.

\section{Conclusions}
\label{conclusion}
In this research, we work on a DM model with two gauge singlet fermionic WIMPs interacting with
the SM particles via a Higgs portal. In fact, a singlet scalar mediator mixes with
the SM Higgs and the interaction of the dark sector with the SM particles becomes possible.
The light partner is stable and is the DM candidate and the heavy partner is a
short-lived particle without contribution in the relic density.

It is found that the singlet fermionic DM model is excluded by the recent direct
detection experiments, except in the resonance region.
This motivates us to extend the minimal singlet DM model to incorporate two WIMPs instead of one.
The DM relic density in our model involves contributions from annihilation and coannihilation processes.
It turns out the coannihilation effects enhance the relic density.

We realized that even in the regions of the parameter space with suppressed coannihilation effects, the
model shows some characteristic features when imposing both DM relic density constraints
and direct detection bounds. In fact we found that there exist regions of the parameter space beyond
the resonance region which evade direct detection upper limits.
These new features arise from a Feynman diagram in the DM annihilation processes with an intermediate 
heavy WIMP which comes in due to its coupling to the DM candidate.
We find that the size of the viable region depends mainly on the singlet scalar mass and the coannihilation effects. 

Moreover, the constraints from Fermi-LAT upper limits on the DM annihilation into $W^+ W^-$ 
and $b \bar b$ are imposed on the viable parameter space. It turned out that only constraints from 
$b \bar b$ channel are strong enough, and can exclude a small portion of the constrained parameter space which 
escape the direct detection experiments.

Further works in this model suggest the search for the mono-X signatures in association
with dark matter pair production at the LHC, for recent reviews see \cite{Kahlhoefer:2017dnp,Penning:2017tmb,Plehn:2017fdg}.
In this direction, a detailed study on various simplified DM models compares the reach of the mono-X searches and direct
searches for the dark mediating particle \cite{Liew:2016oon}.

\section{Acknowledgments}
The author wishes to thank the CERN theoretical physics department for the hospitality and support where
this work was finalized during the visit. Arak University is acknowledged for financial support
under contract no.1397/4.

\label{Ack}

\section{Appendix: Annihilation cross sections}
\label{Apen}
In this section we provide the DM annihilation cross sections with two particles in the final state. 
The cross section with SM fermions in the final state is
\ba
  \sigma v_{rel} (\bar \psi_1 \psi_1 \to \bar f f) = \frac{N_c m^2_f g^2_1 \sin^2 2\theta}{8\pi s v^2_h}
  \left[ (p_1.p_2)^2-2(p_1.p_2) m^2_f +2m^2_f m^2_1 -m^4_1  \right] \times
  \nonumber\\&&\hspace{-11.1cm}
  \Big|\frac{1}{s-m^2_s+im_s \Gamma_s} -\frac{1}{s-m^2_h+im_h \Gamma_h}\Big|^2 \,,
  \ea
where $N_c$ is the number of color.
DM annihilation to a pair of Z boson is

\ba
\sigma v_{rel} (\bar \psi_1 \psi_1 \to ZZ) = \frac{g^2_1 \sin^2 2\theta}{16\pi s v^2_h} 
\Big[ (p_1.p_2)^3-2(p_1.p_2)^2 m^2_Z +(p_1.p_2)^2 m^2_1 + 2(p_1.p_2)m^4_Z
\nonumber\\  
  -(p_1.p_2)m^4_1 -3m^4_Z m^2_1 + 2m^2_Z m^4_1 -m^6_1  \Big] \times
\Big|\frac{1}{s-m^2_s+im_s \Gamma_s} -\frac{1}{s-m^2_h+im_h \Gamma_h}\Big|^2 \,,
\ea
and to a pair of W boson is

\ba
\sigma v_{rel} (\bar \psi_1 \psi_1 \to W^+ W^-) = \frac{g^2_1 \sin^2 2\theta}{8\pi s v^2_h} 
\Big[ (p_1.p_2)^3-2(p_1.p_2)^2 m^2_W +(p_1.p_2)^2 m^2_1 + 2(p_1.p_2)m^4_W
\nonumber\\  
  -(p_1.p_2)m^4_1 -3m^4_W m^2_1 + 2m^2_W m^4_1 -m^6_1  \Big] \times
\Big|\frac{1}{s-m^2_s+im_s \Gamma_s} -\frac{1}{s-m^2_h+im_h \Gamma_h}\Big|^2 \,.
\ea
DM annihilation to a pair of singlet scalar is given by

\ba
\sigma v_{rel} (\bar \psi_1 \psi_1 \to s s) = \frac{\sqrt{1-4m^2_h/s}}{32\pi^2s}
\int d\Omega  \Big[\frac{1}{8} b^2_1 \frac{g^2_1 \sin^2\theta  [p_1.p_2-m^2_1]}{(s-m^2_h)^2}
  + \frac{3}{4} b b_1 \frac{g^2_1 \sin^2 \theta \cos \theta [p_1.p_2-m^2_1]}{(s-m^2_s)(s-m^2_h)}
\nonumber\\ &&\hspace{-17.2cm}
-3 b \frac{g_1 g^2_{12} \sin \theta \cos^3 \theta [2(p_1.p_3)m_1-(p_1.p_2)m_2 -2(p_1.p_2)m_1 +m_2m^2_1]}{(s-m^2_s)(u-m^2_2)} \nonumber\\
+ \frac{9}{8} b^2 \frac{g^2_1 \cos^2\theta \sin^2\theta [p_1.p_2-m^2_1]}{(s-m^2_s)^2} 
+ b_1 \frac{g_1 g^2_{12} \sin \theta \cos^2 \theta [2(p_1.p_3)m_1 +(p_1.p_2)m_2 -m_2 m^2_1 -2 m^3_1]}{(s-m^2_h)(t-m^2_2)} \nonumber\\
+ b_1 \frac{g^3_1 m_1 \sin \theta \cos^2 \theta [2(p_1.p_3) +(p_1.p_2) -3 m^2_1]}{(s-m^2_h)(t-m^2_1)} 
+ 3 b \frac{g^3_1 m_1 \sin \theta \cos^3 \theta [2(p_1.p_3) +(p_1.p_2) -3 m^2_1]}{(s-m^2_s)(t-m^2_1)} \nonumber\\&&\hspace{-17.2cm}
- g^4_1 \cos^4 \theta \frac{[2(p_1.p_3)^2-2(p_1.p_3)(p_1.p_2) -10(p_1.p_3)m^2_1 + (p_1.p_2)m^2_s + m^2_s m^2_1 +8m^4_1]}{(t-m^2_1)^2} \nonumber\\
+ g^4_1 \cos^4 \theta \frac{[2(p_1.p_3)^2-2(p_1.p_3)(p_1.p_2) -2(p_1.p_3)m^2_1 + (p_1.p_2)m^2_s +4(p_1.p_2)m^2_1 + m^2_s m^2_1 -4m^4_1]}
{(t-m^2_1) (u-m^2_1)} \nonumber\\&&\hspace{-17.3cm}
-2 \frac{g^2_1 g^2_{12} \cos^4 \theta}{(t-m^2_1) (t-m^2_2)}  [2(p_1.p_3)^2-2(p_1.p_3)(p_1.p_2) -2(p_1.p_3)m_1 m_2 -8(p_1.p_3)m^2_1 + (p_1.p_2)m^2_s
  \nonumber\\&&\hspace{-13.5cm}
  -(p_1.p_2)m_1 m_2 +(p_1.p_2)m^2_1 + m^2_s m^2_1 +3m_2 m^3_1 + 5 m^4_1] \nonumber\\&&\hspace{-17.2cm}
+2 \frac{g^2_1 g^2_{12} \cos^4 \theta}{(t-m^2_1) (u-m^2_2)}  [2(p_1.p_3)^2-2(p_1.p_3)(p_1.p_2) +2(p_1.p_3)m_1 m_2 -4(p_1.p_3)m^2_1 + (p_1.p_2)m^2_s
  \nonumber\\&&\hspace{-13.5cm}
  +(p_1.p_2)m_1 m_2 +3 (p_1.p_2)m^2_1 + m^2_s m^2_1 -3 m_2 m^3_1 - m^4_1] \nonumber\\&&\hspace{-17.2cm}
- \frac{g^4_{12} \cos^4 \theta}{(t-m^2_2)^2}  [2(p_1.p_3)^2-2(p_1.p_3)(p_1.p_2)-4(p_1.p_3)m_1 m_2 -6(p_1.p_3)m^2_1 + (p_1.p_2)m^2_s \nonumber\\&&\hspace{-15cm}
  -(p_1.p_2)m^2_2 + (p_1.p_2)m^2_1 + m^2_s m^2_1 + m^2_1 m^2_2 +4 m_2 m^3_1 +3 m^4_1] \nonumber\\&&\hspace{-17.2cm}
+ \frac{g^4_{12} \cos^4 \theta}{(t-m^2_2)(u-m^2_2)}  [2(p_1.p_3)^2-2(p_1.p_3)(p_1.p_2)-2(p_1.p_3)m^2_1 + (p_1.p_2) m^2_s \nonumber\\&&\hspace{-15cm}
  +(p_1.p_2)m^2_2 +2 (p_1.p_2)m_1 m_2 +(p_1.p_2)m^2_1 + m^2_s m^2_1 - m^2_1 m^2_2 -2 m_2 m^3_1 - m^4_1] 
\Big] \,,
\ea

where
\ba
b = \sin \theta \cos \theta \lambda_1 - 2 \cos^2 \theta \lambda_2 v_h - \sin^2 \theta \lambda_H v_h \,, 
\nonumber\\&&\hspace{-8.8cm}
b_1 = 3 \sin^3 \theta \lambda_1 - 2 \sin \theta \lambda_1 -6 \cos \theta \sin^2 \theta \lambda_2 v_h + 2 \cos \theta \lambda_2 v_h
  + 3 \cos \theta \sin^2 \theta \lambda_H v_h \,.
\ea
The rerevant annihilation cross section to a pair of Higgs boson is

\ba
\sigma v_{rel} (\bar \psi_1 \psi_1 \to h h) = \frac{\sqrt{1-4m^2_h/s}}{32\pi^2s}
\int d\Omega  \Big[\frac{1}{8} b^2_2 \frac{g^2_1 \cos^2\theta  [p_1.p_2-m^2_1]}{(s-m^2_s)^2}
  -\frac{3}{4}b_3 b_2 \frac{g^2_1 \cos^2 \theta \sin \theta [p_1.p_2-m^2_1]}{(s-m^2_s)(s-m^2_h)}
\nonumber\\&&\hspace{-17.2cm}
+ b_2 m_1 \frac{g^3_1 \cos \theta \sin^2 \theta [2p_1.p_3-3p_1.p_2 +m^2_1]}{(s-m^2_s)(u-m^2_1)} \nonumber\\&&\hspace{-17.2cm}
+ b_2 \frac{g_1 g^2_{12} \cos \theta \sin^2 \theta [2(p_1.p_3)m_1-(p_1.p_2)m_2 -2(p_1.p_2)m_1 +m_2m^2_1]}{(s-m^2_s)(u-m^2_2)}
\nonumber \\&&\hspace{-17.2cm}
+ \frac{9}{8} b^2_3 \frac{g^2_1 \cos^2\theta \sin^2\theta [p_1.p_2-m^2_1]}{(s-m^2_h)^2} 
-3 b_3 m_1 \frac{g^3_1 \cos \theta \sin^3 \theta [2p_1.p_3-3p_1.p_2 +m^2_1]}{(s-m^2_h)(u-m^2_1)} \nonumber\\&&\hspace{-17.2cm}
-3 b_3 \frac{g_1 g^2_{12} \cos \theta \sin^3 \theta [2(p_1.p_3)m_1-(p_1.p_2)m_2 -2(p_1.p_2)m_1 +m_2m^2_1]}{(s-m^2_h)(u-m^2_2)} \nonumber\\&&\hspace{-17.2cm}
- g^4_1 \sin^4 \theta \frac{[2(p_1.p_3)^2-2(p_1.p_3)(p_1.p_2) -10(p_1.p_3)m^2_1 + (p_1.p_2)m^2_h + m^2_h m^2_1 +8m^4_1]}{(t-m^2_1)^2} \nonumber\\
+ g^4_1 \sin^4 \theta \frac{[2(p_1.p_3)^2-2(p_1.p_3)(p_1.p_2) -2(p_1.p_3)m^2_1 + (p_1.p_2)m^2_h +4(p_1.p_2)m^2_1 + m^2_h m^2_1 -4m^4_1]}
{(t-m^2_1) (u-m^2_1)} \nonumber\\&&\hspace{-17.2cm}
-2 \frac{g^2_1 g^2_{12} \sin^4 \theta}{(t-m^2_1) (t-m^2_2)}  [2(p_1.p_3)^2-2(p_1.p_3)(p_1.p_2) -2(p_1.p_3)m_1 m_2 -8(p_1.p_3)m^2_1 + (p_1.p_2)m^2_h
  \nonumber\\&&\hspace{-13.6cm}
  -(p_1.p_2)m_1 m_2 +(p_1.p_2)m^2_1 + m^2_h m^2_1 +3m_2 m^3_1 + 5 m^4_1] \nonumber\\&&\hspace{-17.2cm}
+2 \frac{g^2_1 g^2_{12} \sin^4 \theta}{(t-m^2_1) (u-m^2_2)}  [2(p_1.p_3)^2-2(p_1.p_3)(p_1.p_2) +2(p_1.p_3)m_1 m_2 -4(p_1.p_3)m^2_1 + (p_1.p_2)m^2_h
  \nonumber\\&&\hspace{-13.6cm}
  +(p_1.p_2)m_1 m_2 +3 (p_1.p_2)m^2_1 + m^2_h m^2_1 -3 m_2 m^3_1 - m^4_1] \nonumber\\&&\hspace{-17.2cm}
- \frac{g^4_{12} \sin^4 \theta}{(t-m^2_2)^2}  [2(p_1.p_3)^2-2(p_1.p_3)(p_1.p_2)-4(p_1.p_3)m_1 m_2 -6(p_1.p_3)m^2_1 + (p_1.p_2)m^2_h
  \nonumber\\&&\hspace{-13.6cm}
  -(p_1.p_2)m^2_2 + (p_1.p_2)m^2_1 + m^2_h m^2_1 + m^2_1 m^2_2 +4 m_2 m^3_1 +3 m^4_1] \nonumber\\&&\hspace{-17.2cm}
+ \frac{g^4_{12} \sin^4 \theta}{(t-m^2_2)(u-m^2_2)}  [2(p_1.p_3)^2-2(p_1.p_3)(p_1.p_2)-2(p_1.p_3)m^2_1 + (p_1.p_2)m^2_h \nonumber\\
  +(p_1.p_2)m^2_2 +2 (p_1.p_2)m_1 m_2 +(p_1.p_2)m^2_1 + m^2_h m^2_1 - m^2_1 m^2_2 -2 m_2 m^3_1 - m^4_1] 
\Big]\,,
\ea

where
\ba
b_2 = 3  \lambda_1 \cos \theta \sin^2 \theta  - \lambda_1 \cos \theta  + 6 \lambda_2 v_h \sin^3 \theta
      -4 \lambda_2 v_h \sin \theta + 3 \lambda_H v_h \cos^2 \theta \sin \theta  \,, 
\nonumber\\&&\hspace{-14.7cm}
b_3 = \lambda_1 \cos \theta \sin \theta + 2 \lambda_2 v_h \sin^2 \theta + \lambda_H v_h \cos^2 \theta  \,.
\ea
Finally, the DM annihilation to a singlet scalar and SM Higgs boson is 
\ba
\sigma v_{rel} (\bar \psi_1 \psi_1 \to h s) = \frac{\sqrt{1-4m^2_h/s}}{32\pi^2s}
\int d\Omega  \Big[\frac{1}{4} b^2_1 \frac{g^2_1 \cos^2\theta  [p_1.p_2-m^2_1]}{(s-m^2_s)^2}
   + \frac{1}{4} b^2_2 \frac{g^2_1 \sin^2\theta  [p_1.p_2-m^2_1]}{(s-m^2_h)^2}
  \nonumber\\&&\hspace{-16cm}
  -\frac{1}{2}b_1 b_2 \frac{g^2_1 \cos \theta \sin \theta [p_1.p_2-m^2_1]}{(s-m^2_s)(s-m^2_h)}
+ \frac{1}{2} b_1 m_1 \frac{g^3_1 \cos^2 \theta \sin \theta [4 p_1.p_3 + 2 p_1.p_2 + m^2_h -6m^2_1 -m^2_s]}{(s-m^2_s)(t-m^2_1)}
\nonumber\\&&\hspace{-16cm}
- \frac{1}{2} b_2 m_1 \frac{g^3_1 \cos \theta \sin^2 \theta [4 p_1.p_3 + 2 p_1.p_2 + m^2_h -6m^2_1 -m^2_s]}{(s-m^2_h)(t-m^2_1)}
\nonumber\\&&\hspace{-16cm}
- \frac{1}{2} b_1 m_1 \frac{g^3_1 \cos^2 \theta \sin \theta [4 p_1.p_3 - 6 p_1.p_2 + m^2_h +2 m^2_1 - m^2_s]}{(s-m^2_s)(u-m^2_1)}
\nonumber\\&&\hspace{-16cm}
+ \frac{1}{2} b_2 m_1 \frac{g^3_1 \cos \theta \sin^2 \theta [4 p_1.p_3 - 6 p_1.p_2 + m^2_h +2 m^2_1 - m^2_s]}{(s-m^2_h)(u-m^2_1)}
\nonumber\\&&\hspace{-16cm}
+ \frac{1}{2} b_1 \frac{g_1 g^2_{12} \cos^2 \theta \sin \theta [4(p_1.p_3)m_1 +2(p_1.p_2)m_2 +m^2_h m_1 -2m^2_1 m_2 - 4 m^3_1 - m_1 m^2_s]}{(s-m^2_s)(t-m^2_2)}
\nonumber\\&&\hspace{-16cm}
- \frac{1}{2} b_2 \frac{g_1 g^2_{12} \cos \theta \sin^2 \theta [4(p_1.p_3)m_1 +2(p_1.p_2)m_2 +m^2_h m_1 -2m^2_1 m_2 - 4 m^3_1 - m_1 m^2_s]}{(s-m^2_h)(t-m^2_2)}
\nonumber\\&&\hspace{-16cm}
+ \frac{1}{2} b_2 \frac{g_1 g^2_{12} \cos \theta \sin^2 \theta [4(p_1.p_3)m_1 -2(p_1.p_2)m_2  - 4 (p_1.p_2)m_1 + m^2_h m_1 +2m^2_1 m_2 - m_1 m^2_s]}{(s-m^2_h)(u-m^2_2)}
\nonumber\\&&\hspace{-16cm}
- \frac{g^4_{1} \sin^2 \theta \cos^2 \theta}{(t-m^2_1)^2}  [2(p_1.p_3)^2-2(p_1.p_3)(p_1.p_2)-10(p_1.p_3)m^2_1 + (p_1.p_3)m^2_h -(p_1.p_3)m^2_s
\nonumber\\&&\hspace{-13cm}
   +(p_1.p_2)m^2_s -2 m^2_h m^2_1 +8 m^4_1 + 3 m^2_1 m^2_s]
\nonumber\\&&\hspace{-16cm}
- \frac{g^4_{1} \sin^2 \theta \cos^2 \theta}{(u-m^2_1)^2}  [2(p_1.p_3)^2-2(p_1.p_3)(p_1.p_2) +6 (p_1.p_3)m^2_1 + (p_1.p_3)m^2_h  -(p_1.p_3)m^2_s
  \nonumber\\&&\hspace{-13cm}
  +(p_1.p_2)m^2_s  -8 (p_1.p_2)m^2_1 +2 m^2_h m^2_1 - m^2_1 m^2_s]
\nonumber\\&&\hspace{-16cm}
+ 2\frac{ g^4_{1} \sin^2 \theta \cos^2 \theta}{(t-m^2_1)(u-m^2_1)}  [2(p_1.p_3)^2-2(p_1.p_3)(p_1.p_2) -2 (p_1.p_3)m^2_1 + (p_1.p_3)m^2_h
  \nonumber\\&&\hspace{-13cm}
  -(p_1.p_3)m^2_s +(p_1.p_2)m^2_s  +4 (p_1.p_2)m^2_1  - 4 m^4_1 +  m^2_1 m^2_s]
\nonumber\\&&\hspace{-16cm}
+ \frac{ g^2_{1} g^2_{12} \sin^2 \theta \cos^2 \theta}{(t-m^2_2)(u-m^2_1)}  [4(p_1.p_3)^2 -4(p_1.p_3)(p_1.p_2) -4 (p_1.p_3)m_1 m_2 + 2(p_1.p_3)m^2_h 
  \nonumber\\&&\hspace{-13cm}
  -2(p_1.p_3)m^2_s +2(p_1.p_2)m^2_s  +6 (p_1.p_2)m_1 m_2 +2 (p_1.p_2)m^2_1 - m^2_h m_1 m_2
  \nonumber\\&&\hspace{-13cm}
  + m^2_h m^2_1 - 2 m_2 m^3_1 + m_2 m_1 m^2_s - 6m^4_1 + m^2_1 m^2_s]
\nonumber\\&&\hspace{-16cm}
- \frac{ g^2_{1} g^2_{12} \sin^2 \theta \cos^2 \theta}{(t-m^2_1)(t-m^2_2)}  [4(p_1.p_3)^2 -4(p_1.p_3)(p_1.p_2) -4 (p_1.p_3)m_1 m_2 + 2(p_1.p_3)m^2_h 
  -16(p_1.p_3)m^2_1
\nonumber\\&&\hspace{-13cm}
 -2(p_1.p_3)m^2_s  -2 (p_1.p_2) m_1 m_2 +2 (p_1.p_2)m^2_1 +2 (p_1.p_2)m^2_s - m^2_h m_1 m_2
\nonumber\\&&\hspace{-13cm}
  -3 m^2_h m^2_1 +6 m_2 m^3_1 + m_2 m_1 m^2_s +10 m^4_1 +5 m^2_1 m^2_s]
\nonumber\\&&\hspace{-16cm}
+ \frac{ g^2_{1} g^2_{12} \sin^2 \theta \cos^2 \theta}{(t-m^2_1)(u-m^2_2)}  [4(p_1.p_3)^2 -4(p_1.p_3)(p_1.p_2) +4 (p_1.p_3)m_1 m_2 + 2(p_1.p_3)m^2_h 
   \nonumber\\&&\hspace{-13cm}
   -8(p_1.p_3)m^2_1 -2(p_1.p_3)m^2_s  +2 (p_1.p_2) m_1 m_2 +6(p_1.p_2)m^2_1 +2 (p_1.p_2)m^2_s
   \nonumber\\&&\hspace{-13cm}
   + m^2_h m_1 m_2  - m^2_h m^2_1 -6 m_2 m^3_1 - m_2 m_1 m^2_s -2 m^4_1 + 3 m^2_1 m^2_s]
\nonumber\\&&\hspace{-16cm}
- \frac{ g^2_{1} g^2_{12} \sin^2 \theta \cos^2 \theta}{(u-m^2_1)(u-m^2_2)}  [4(p_1.p_3)^2 -4(p_1.p_3)(p_1.p_2) +4 (p_1.p_3)m_1 m_2 + 2(p_1.p_3)m^2_h 
   \nonumber\\&&\hspace{-13cm}
   +8(p_1.p_3)m^2_1 -2(p_1.p_3)m^2_s -6 (p_1.p_2) m_1 m_2 -10(p_1.p_2)m^2_1 +2 (p_1.p_2)m^2_s
   \nonumber\\&&\hspace{-13cm}
   +m^2_h m_1 m_2  +3 m^2_h m^2_1 +2 m_2 m^3_1 - m_2 m_1 m^2_s -2 m^4_1 - m^2_1 m^2_s]
\nonumber\\&&\hspace{-16cm}
- \frac{ g^4_{12} \sin^2 \theta \cos^2 \theta}{(t-m^2_2)^2} [2(p_1.p_3)^2 -2(p_1.p_3)(p_1.p_2) - 4 (p_1.p_3)m_1 m_2 + (p_1.p_3)m^2_h 
  \nonumber\\&&\hspace{-13cm}
  -6 (p_1.p_3) m^2_1 - (p_1.p_3) m^2_s  - (p_1.p_3)m^2_s - (p_1.p_2)m^2_2 + (p_1.p_2)m^2_1
  \nonumber\\&&\hspace{-13cm}
  + (p_1.p_2)m^2_s - m^2_h m_1 m_2  - m^2_h m^2_1 + m^2_2 m^2_1
  +4 m_2 m^3_1 + m_2 m_1 m^2_s
\nonumber\\&&\hspace{-13cm}
  +3 m^4_1 +2 m^2_1 m^2_s]
\nonumber\\&&\hspace{-16cm}
+2 \frac{ g^4_{12} \sin^2 \theta \cos^2 \theta}{(t-m^2_2)(u-m^2_2)} [2(p_1.p_3)^2 -2(p_1.p_3)(p_1.p_2)  + (p_1.p_3)m^2_h 
  -2 (p_1.p_3) m^2_1
\nonumber\\&&\hspace{-13cm}
  - (p_1.p_3) m^2_s  + (p_1.p_2)m^2_s + (p_1.p_2) m^2_2 + (p_1.p_2) m^2_1 + 2(p_1.p_2) m_1 m_2
  \nonumber\\&&\hspace{-13cm}
  - m^2_2 m^2_1 -2 m_2 m^3_1 - m^4_1 + m^2_1 m^2_s]
\nonumber\\&&\hspace{-16cm}
- \frac{ g^4_{12} \sin^2 \theta \cos^2 \theta}{(u-m^2_2)^2} [2(p_1.p_3)^2 -2(p_1.p_3)(p_1.p_2) + 4 (p_1.p_3)m_1 m_2
  - 4 (p_1.p_2)m_1 m_2
\nonumber\\&&\hspace{-13cm}
+ (p_1.p_3)m^2_h +2 (p_1.p_3) m^2_1 - (p_1.p_3) m^2_s + (p_1.p_2)m^2_s - (p_1.p_2)m^2_2
\nonumber\\&&\hspace{-13cm}
-3 (p_1.p_2)m^2_1 +  m^2_h m_1 m_2  + m^2_h m^2_1 + m^2_2 m^2_1
- m_2 m_1 m^2_s - m^4_1 ]
\nonumber\\
\Big] \,.
\ea

\bibliography{ref}
\bibliographystyle{utphys}

\end{document}